\pdfoutput=1
\documentclass[12pt]{article}

\usepackage[margin=1in]{geometry}
\usepackage{amsmath,amssymb,amsthm}
\usepackage[round]{natbib}
\usepackage{setspace}
\usepackage{enumitem}
\usepackage{placeins}
\usepackage{xcolor}

\usepackage[hidelinks]{hyperref}
\hypersetup{pdftitle={On the difficulty of characterizing top trading cycles with identical houses},pdfauthor={Bettina Klaus}}

\newtheoremstyle{boldplain}{\topsep}{\topsep}{\itshape}{0pt}{\bfseries}{.}{5pt plus 1pt minus 1pt}{\thmname{#1}\thmnumber{ #2}\thmnote{ (#3)}}
\newtheoremstyle{bolddef}{\topsep}{\topsep}{\normalfont}{0pt}{\bfseries}{.}{5pt plus 1pt minus 1pt}{\thmname{#1}\thmnumber{ #2}\thmnote{ (#3)}}
\theoremstyle{boldplain}
\newtheorem{theorem}{Theorem}
\newtheorem{proposition}{Proposition}
\newtheorem{lemma}{Lemma}

\newtheorem{citedtheorem}{Theorem}

\theoremstyle{bolddef}
\newtheorem{definition}{Definition}
\theoremstyle{definition}
\newtheorem{remark}{Remark}
\makeatletter
\renewenvironment{proof}[1][\proofname]{\par\pushQED{\qed}\normalfont\topsep6\p@\@plus6\p@\relax\trivlist\item[\hskip\labelsep\bfseries#1\@addpunct{.}]\ignorespaces}{\popQED\endtrivlist\@endpefalse}
\makeatother

\newcommand{\Nc}{\mathcal{N}}
\newcommand{\Rd}{\mathcal{R}}
\newcommand{\T}{T}

\title{On the difficulty of characterizing top trading cycles\\ with identical houses}
\author{Bettina Klaus\thanks{Faculty of Business and Economics, University of Lausanne, Internef 538, CH-1015 Lausanne, Switzerland; e-mail: \href{mailto:bettina.klaus@unil.ch}{bettina.klaus@unil.ch}. I thank Jacob Coreno, Di Feng, and Mehmet Karakaya for helpful comments and suggestions. I gratefully acknowledge support from the Swiss National Science Foundation (SNSF) through Project 100018-231777. During the preparation of this work, I used Claude (Anthropic) and ChatGPT (OpenAI) to check proofs and examples ex post and to verify the minimality result of Appendix~\ref{app:search}.}}
\date{\today}

\begin{document}

\maketitle

\begin{abstract}
We study Shapley--Scarf housing markets (Shapley and Scarf, 1974) in which houses come in types and agents are indifferent between houses of the same type but rank the types strictly. On this domain of objective indifferences, top trading cycles (TTC) rules with fixed tie-breaking regain \textsl{Pareto efficiency} and \textsl{group strategy-proofness} (Sandholtz and Tai, 2026), properties that need not hold under general weak preferences. We ask whether the characterization of the TTC rule for strict preferences by Ma (1994) and the characterization of TTC rules with fixed tie-breaking for weak preferences by Ehlers (2014) extend to this domain. They do not. With five agents and two types of houses, we construct a rule that satisfies \textsl{individual rationality}, \textsl{Pareto efficiency}, \textsl{group strategy-proofness}, \textsl{non-bossiness}, and \textsl{consistency}, but whose restriction to the five-agent population is not welfare equivalent to any TTC rule with fixed tie-breaking. With two types, five agents are minimal for this fixed-population result. The same rule shows that several other characterizations of TTC for strict preferences do not extend either.

\end{abstract}

\section{Introduction}

Housing markets \citep{ShapleyScarf1974} and the top trading cycles (TTC) rule, which \citet{ShapleyScarf1974} attribute to David Gale, have been studied extensively; \citet{AfacanHuLi2024} survey the literature on housing markets since \citet{ShapleyScarf1974}, and \citet{MorrillRoth2024} survey TTC and its applications. For housing markets with strict preferences, the TTC rule is \textsl{individually rational} and \textsl{Pareto efficient} \citep{ShapleyScarf1974,RothPostlewaite1977}, \textsl{strategy-proof} \citep{Roth1982}, and \textsl{group strategy-proof} \citep{Bird1984,SandholtzTai2024}. Moreover, for strict preferences, \citet{Ma1994} shows that the TTC rule is the only rule satisfying \textsl{individual rationality}, \textsl{Pareto efficiency}, and \textsl{strategy-proofness}. Several authors refine this characterization by weakening \textsl{Pareto efficiency} or \textsl{strategy-proofness} or by replacing them with other properties \citep{Takamiya2001,FujinakaWakayama2018,AltuntasPhanTamura2023,Ekici2024,CorenoFeng2025,CorenoFeng2026,ChenJiaoZhangZhang2025}. For general weak preferences, TTC rules with fixed tie-breaking, which break ties in the agents' preferences according to exogenously fixed orders before applying TTC, need not satisfy \textsl{Pareto efficiency} \citep{JaramilloManjunath2012,Ehlers2014} or \textsl{group strategy-proofness} \citep{SandholtzTai2026}. \citet{Ehlers2014} characterizes TTC rules with fixed tie-breaking, up to welfare equivalence (two rules are welfare equivalent if each agent is always indifferent between his allotments under the two rules), by \textsl{individual rationality}, \textsl{weak Pareto efficiency}, \textsl{strategy-proofness}, \textsl{non-bossiness}, and \textsl{consistency}; when indifference with an agent's endowment is allowed, each agent has two fixed tie-breaking orders, one for ties among houses he finds indifferent to his endowment and one for all other ties.

\citet{SandholtzTai2026} study housing markets in which houses come in types, houses of the same type are identical, and each agent is indifferent between the houses of a type and ranks the types strictly. As they point out, such objective indifferences arise naturally when several available objects are effectively indistinguishable from the agents' point of view, for example identical dorm rooms or school seats. On this domain of objective indifferences, which is a subdomain of the weak-preference domain and contains the strict-preference domain as the extreme case in which each type consists of a single house, \citet{SandholtzTai2026} show that TTC rules with fixed tie-breaking regain \textsl{Pareto efficiency} and \textsl{group strategy-proofness}. They further show that objective indifferences are, in a precise sense, the most general indifferences for which TTC rules with fixed tie-breaking retain these properties. They explicitly leave an axiomatic characterization of TTC rules with fixed tie-breaking on this domain for future work. A natural first step is to ask whether the characterizations of \citet{Ma1994} and \citet{Ehlers2014}, the benchmark results for strict preferences and for weak preferences, respectively, extend to this domain.

We show that neither the characterization of \citet{Ma1994} nor that of \citet{Ehlers2014} extends to the domain of objective indifferences. For a population with three agents owning houses of one type and two agents owning houses of the other type, we construct a rule that satisfies \textsl{individual rationality}, \textsl{Pareto efficiency}, \textsl{group strategy-proofness}, \textsl{non-bossiness}, and \textsl{consistency}, but that is not welfare equivalent to any TTC rule with fixed tie-breaking. Since \textsl{consistency} relates problems with different sets of agents, \citet{Ehlers2014} considers rules defined for all coalitions of a finite set of potential agents, whereas \citet{Ma1994} and most refinements of his characterization consider rules for a fixed set of agents. Our rule covers both settings: it is defined for all coalitions, and even on the problems with all five agents it is not welfare equivalent to any TTC rule with fixed tie-breaking.

Two observations explain why these characterizations do not extend. First, the set of possible reports is much smaller than for strict or general weak preferences: each agent can only report a ranking of the types. In particular, an agent cannot rank a house above an identical house; hence, whenever the house he receives or his own house has an identical copy, he cannot rank the house he receives first and his own house second. Proofs of the characterization of \citet{Ma1994} rely on auxiliary profiles that contain such reports. For restricted domains of strict preferences, \citet{GoelTamura2026} show that this characterization extends whenever their top-two condition guarantees these reports, and that it fails on many domains that violate the condition, including all such domains with at most four houses. Our counterexample has a structure similar to that of their non-TTC rules: it coincides with a TTC rule except at a profile at which these reports are unavailable, and there one agent receives a house of his second-ranked type instead of his TTC allotment. Our result concerns objective indifferences rather than restricted domains of strict preferences and, more importantly, shows that the characterization fails even when \textsl{group strategy-proofness}, \textsl{non-bossiness}, and \textsl{consistency} are also imposed. Second, \textsl{consistency} can remove only groups whose collective allotments are their collective endowments; our formulation is the same as that of \citet{Ehlers2014}.\footnote{\citet{KarakayaKlaus2026} study a type-based formulation of \textsl{consistency}, which also admits departures of groups that do not leave with their own endowments; see Section~\ref{sec:robust}.} In the proof of \citet{Ehlers2014}, \textsl{consistency} works together with strict reports: an agent's preferences are replaced by a strict preference that leaves the outcome unchanged, or by one that ranks a particular house first and his own house second, and only then does \textsl{consistency} remove the agents who keep their endowments or trade among themselves. Since his proof requires these strict reports, his characterization extends to every domain that contains all strict preferences \citep[Theorem~3]{Ehlers2014}, but his proof does not apply to our domain, which does not contain them. In our counterexample, one of the five agents keeps his house and the other four agents trade in one cycle. Thus, \textsl{consistency} can remove either this agent or the four agents of the cycle together, but no proper subset of the four agents, and therefore relates their trades to no problem with fewer of them. Consequently, the properties cannot force the rule to break ties within a type in the same way at all preference profiles, as a fixed tie-breaking structure does.

The counterexample has consequences beyond the characterizations of \citet{Ma1994} and \citet{Ehlers2014}. Several closely related characterizations of TTC for strict preferences \citep{Takamiya2001,AltuntasPhanTamura2023,Ekici2024,EkiciSethuraman2024,CorenoFeng2026,ChenJiaoZhangZhang2025} do not extend to the domain of objective indifferences either: when their properties are formulated on this domain, treating houses of the same type as identical, our rule satisfies them. Since \textsl{endowments-swapping-proofness} \citep{FujinakaWakayama2018} compares problems with different assignments of houses to agents, we extend our rule to variable endowments; the extended rule satisfies \textsl{endowments-swapping-proofness} (Proposition~\ref{prop:esp}), and hence the characterizations of TTC that use this property do not extend either.

With two types of houses, five agents are the smallest number of agents for which a fixed-population counterexample exists; we establish this minimality by a computer search (Appendix~\ref{app:search}). What matters is not only the number of agents but the type composition: with five agents, a fixed-population counterexample exists if three agents own houses of one type and two agents own houses of the other type, but not if four agents own houses of one type and a single agent owns a house of the other type. In Appendix~\ref{app:four}, we show that a counterexample to extending the characterization of \citet{Ehlers2014} already exists with four agents. This four-agent counterexample does not apply to the characterization of \citet{Ma1994}: on the problems with all four agents, the rule is welfare equivalent to a TTC rule with fixed tie-breaking, and its failure of welfare equivalence arises only through its behavior on a problem with three agents; with two types, four agents are minimal for such a counterexample.

Thus, although TTC rules with fixed tie-breaking regain \textsl{Pareto efficiency} and \textsl{group strategy-proofness} on the domain of objective indifferences, the properties that characterize the TTC rule for strict preferences and TTC rules with fixed tie-breaking for weak preferences do not single out TTC rules with fixed tie-breaking on this domain. A characterization of these rules therefore requires properties beyond those used in the characterizations that we consider.

The paper is organized as follows. In Section~\ref{sec:model}, we introduce the model and the properties, establish the relation between \textsl{group strategy-proofness} and \textsl{welfare non-bossiness} (the proof is given in Appendix~\ref{app:gsp}), introduce TTC rules with fixed tie-breaking, and state the characterizations of \citet{Ma1994} and \citet{Ehlers2014}. In Section~\ref{sec:cex}, we present the main counterexample. In Section~\ref{sec:robust}, we show that several other characterizations of TTC for strict preferences do not extend either, and we conclude. Appendix~\ref{app:four} contains the four-agent counterexample, and in Appendix~\ref{app:search}, we establish the fixed-population minimality result by a computer search.

\section{Model}\label{sec:model}

Let $\Nc$ be a finite set of potential agents and let $\T$ be a finite set of house types. Each agent $i\in\Nc$ owns a house $h(i)$, his \emph{endowment}, of type $\tau(h(i))\in\T$, and different agents own different houses, i.e., $h(i)\neq h(j)$ whenever $i\neq j$. We also refer to $\Nc$ as the \emph{population}. The \emph{type composition} of the population specifies, for each type, the number of agents who own a house of that type. With two types $a$ and $b$, we write it as a sequence of letters, one for each agent's endowment, with all $a$'s first; for instance, type composition $aaabb$ means that three agents own houses of type $a$ and two agents own houses of type $b$. A \emph{coalition} is a non-empty set $N\subseteq\Nc$; let $H_N=\{h(i)\}_{i\in N}$ denote the set of houses of coalition $N$ and let $\T_N=\{\tau(h):h\in H_N\}$ denote the set of types present in $N$.

For each coalition $N$, let $\Rd(H_N)$ denote the set of complete and transitive preferences over $H_N$ that are indifferent between houses of the same type and strict between houses of different types \citep[the domain of objective indifferences of][]{SandholtzTai2026}. For each $R_i\in\Rd(H_N)$, let $P_i$ and $I_i$ denote its asymmetric and symmetric parts. Since preferences in $\Rd(H_N)$ are strict across types, each $R_i\in\Rd(H_N)$ induces a linear order over $\T_N$, which we denote by $R_i$ as well. A pair $(N,R)$ of a coalition $N$ and a preference profile $R=(R_i)_{i\in N}\in\Rd(H_N)^N$ is a \emph{problem}.

A \emph{matching} for coalition $N$ is a bijection $\mu$ from $N$ to $H_N$ that assigns to each $i\in N$ an \emph{allotment} $\mu(i)\in H_N$. A \emph{trading cycle} of $\mu$ is a sequence of distinct agents $i_1,\ldots,i_m\in N$ such that $\mu(i_l)=h(i_{l+1})$ for each $l\in\{1,\ldots,m-1\}$ and $\mu(i_m)=h(i_1)$; we denote it by $i_1\to h(i_2)\to i_2\to\cdots\to i_m\to h(i_1)\to i_1$, i.e., each agent points to his allotment and each house points to its owner. A trading cycle $i_1\to h(i_1)\to i_1$ means that agent $i_1$ keeps his endowment. Each agent in $N$ belongs to exactly one trading cycle of a matching; hence, a group of agents receives precisely its endowments if and only if it consists of all agents of some trading cycles. Since agents are indifferent between houses of the same type and strict between houses of different types, their welfare depends only on the type of the house they receive. Accordingly, two matchings $\mu$ and $\mu'$ for $N$ are \emph{welfare equivalent} if for each $i\in N$, $\tau(\mu(i))=\tau(\mu'(i))$, i.e., if they induce the same \emph{type allocation} $\tau(\mu)=(\tau(\mu(i)))_{i\in N}$, which lists the type of each agent's allotment.

Since \textsl{consistency}, introduced below, compares problems with different coalitions, a rule is defined simultaneously for all coalitions. A \emph{rule} $f$ assigns to each problem $(N,R)$ a matching $f(N,R)$ for $N$; agent $i$'s allotment is $f_i(N,R)$. Two rules $f$ and $f'$ are \emph{welfare equivalent} if for each problem $(N,R)$, $f(N,R)$ and $f'(N,R)$ are welfare equivalent. For each coalition $N$, let $f(N,\cdot)$ denote the restriction of $f$ to the problems with coalition $N$; $f(N,\cdot)$ and $f'(N,\cdot)$ are \emph{welfare equivalent} if for each $R\in\Rd(H_N)^N$, $f(N,R)$ and $f'(N,R)$ are welfare equivalent.

\subsection*{Properties}

We next introduce the properties of rules that we consider.\medskip

The well-known property of \textsl{individual rationality} requires that no agent receives a house that he finds worse than his own house.

\begin{definition}[Individual rationality]
A matching $\mu$ for $N$ is \emph{individually rational} for $(N,R)$ if for each $i\in N$, $\mu(i)\,R_i\,h(i)$. A rule $f$ is \emph{individually rational} if for each problem $(N,R)$, $f(N,R)$ is \textsl{individually rational} for $(N,R)$.
\end{definition}

\textsl{Pareto efficiency} requires that no other matching makes some agent better off without making any agent worse off; \textsl{weak Pareto efficiency} only requires that no other matching makes all agents better off.

\begin{definition}[Pareto efficiency and weak Pareto efficiency]
A matching $\mu$ for $N$ is \emph{Pareto efficient} for $(N,R)$ if there is no matching $\mu'$ for $N$ such that for each $i\in N$, $\mu'(i)\,R_i\,\mu(i)$, and for some $j\in N$, $\mu'(j)\,P_j\,\mu(j)$. A matching $\mu$ for $N$ is \emph{weakly Pareto efficient} for $(N,R)$ if there is no matching $\mu'$ for $N$ such that for each $i\in N$, $\mu'(i)\,P_i\,\mu(i)$. A rule $f$ is \emph{Pareto efficient} (\emph{weakly Pareto efficient}) if for each problem $(N,R)$, $f(N,R)$ is \textsl{Pareto efficient} (\textsl{weakly Pareto efficient}) for $(N,R)$.
\end{definition}

The well-known non-manipulability property \textsl{strategy-proofness} requires that no agent can ever benefit from misrepresenting his preferences.

\begin{definition}[Strategy-proofness]
A rule $f$ is \emph{strategy-proof} if for each problem $(N,R)$, each $i\in N$, and each $R'_i\in\Rd(H_N)$, $f_i(N,R)\,R_i\,f_i(N,(R'_i,R_{-i}))$, i.e., no agent $i$ can manipulate $f$ at $(N,R)$ via $R'_i$.
\end{definition}

\textsl{Group strategy-proofness} strengthens \textsl{strategy-proofness} by requiring that no group of agents can misreport their preferences such that each member of the group weakly gains and some member strictly gains.

\begin{definition}[Group strategy-proofness]
Let $f$ be a rule and $(N,R)$ be a problem. A group $S\subseteq N$ \emph{manipulates} $f$ at $(N,R)$ via reported preferences $R'_S\in\Rd(H_N)^S$ if, for $R'=(R'_S,R_{-S})$, for each $i\in S$, $f_i(N,R')\,R_i\,f_i(N,R)$, and for some $j\in S$, $f_j(N,R')\,P_j\,f_j(N,R)$. The rule $f$ is \emph{group strategy-proof} if for each problem $(N,R)$, no group $S\subseteq N$ manipulates $f$ at $(N,R)$.
\end{definition}

\textsl{Non-bossiness} requires that an agent who does not change his own allotment by changing his reported preferences does not change any other agent's allotment; \textsl{welfare non-bossiness} requires that an agent who does not change his own welfare by changing his reported preferences does not change any other agent's welfare. For strict preferences, the two properties coincide.

\begin{definition}[Non-bossiness and welfare non-bossiness]
A rule $f$ is \emph{non-bossy} if for each problem $(N,R)$, each $i\in N$, and each $R'_i\in\Rd(H_N)$, if $f_i(N,(R'_i,R_{-i}))=f_i(N,R)$, then $f(N,(R'_i,R_{-i}))=f(N,R)$. A rule $f$ is \emph{welfare non-bossy} if for each problem $(N,R)$, each $i\in N$, and each $R'_i\in\Rd(H_N)$, if $f_i(N,(R'_i,R_{-i}))\,I_i\,f_i(N,R)$, then for each $j\in N$, $f_j(N,(R'_i,R_{-i}))\,I_j\,f_j(N,R)$.
\end{definition}

For strict preferences, a rule is \textsl{group strategy-proof} if and only if it is \textsl{strategy-proof} and \textsl{non-bossy} \citep{Papai2000,Alva2017}. On the domain of objective indifferences, this equivalence fails, but if \textsl{non-bossiness} is replaced by \textsl{welfare non-bossiness}, an analogous equivalence holds.

\begin{proposition}[Group strategy-proofness and welfare non-bossiness]\label{prop:gsp}
A rule is \textsl{group strategy-proof} if and only if it is \textsl{strategy-proof} and \textsl{welfare non-bossy}. However, \textsl{group strategy-proofness} does not imply \textsl{non-bossiness}, and \textsl{strategy-proofness} together with \textsl{non-bossiness} does not imply \textsl{group strategy-proofness}.
\end{proposition}

We prove Proposition~\ref{prop:gsp} in Appendix~\ref{app:gsp}. The first statement of Proposition~\ref{prop:gsp} parallels \citet[Proposition~1]{PyciaUnver2025}, who show that, on rich domains, which allow for indifferences, \textsl{group strategy-proofness} is equivalent to \textsl{strategy-proofness} and \textsl{non-bossiness} formulated in terms of welfare.\medskip

Next, we formulate \textsl{consistency}: if some agents leave with their allotments and these allotments are precisely their own endowments, then the rule should allocate the remaining houses among the remaining agents in the same way as in the original problem. Let $(N,R)$ be a problem, let $S\subsetneq N$ be a non-empty group of agents, and let $R|_{N\setminus S}$ denote the preference profile of the agents in $N\setminus S$ restricted to $H_{N\setminus S}$. Removing $S$ together with its endowments $H_S$ yields the \emph{reduced problem} $(N\setminus S,R|_{N\setminus S})$; since each remaining agent still owns his original endowment, $(N\setminus S,R|_{N\setminus S})$ is a problem.

\begin{definition}[Consistency]
A rule $f$ is \emph{consistent} if for each problem $(N,R)$, each non-empty group $S\subsetneq N$ whose allotments are precisely its endowments, i.e., $\{f_i(N,R)\}_{i\in S}=H_S$, and each $j\in N\setminus S$, $f_j(N\setminus S,R|_{N\setminus S})=f_j(N,R)$.
\end{definition}

\subsection*{TTC rules with fixed tie-breaking}

A \emph{tie-breaking structure} $\succ=(\succ_i)_{i\in\Nc}$ consists of strict linear orders $\succ_i$ over $\Nc$. Given $\succ$, a problem $(N,R)$, and an agent $i\in N$, agent $i$'s strict preference $R^\succ_i$ over $H_N$, with asymmetric part $P^\succ_i$, ranks houses of different types according to $R_i$ and houses of the same type according to $\succ_i$, i.e., for all $j,k\in N$ with $\tau(h(j))=\tau(h(k))$ and $j\neq k$, $h(j)\,P^\succ_i\,h(k)$ if and only if $j\succ_i k$. The \emph{TTC rule based on $\succ$}, $\varphi^\succ$, assigns to each problem $(N,R)$ the outcome of the top trading cycles algorithm, which \citet{ShapleyScarf1974} attribute to David Gale \citep[see also][]{Scarf2009}, for $(N,R^\succ)$: at each step, each remaining agent $i$ points to his most preferred remaining house according to $R^\succ_i$, each remaining house points to its owner, and the agents in each cycle receive the houses they point to and are removed together with these houses. Since each house points to its owner, a house is removed only together with its owner; hence, each agent's own house remains as long as the agent remains. Since tie-breaking orders are consulted only to compare houses of the same type, $\varphi^\succ$ depends on $\succ$ only through each agent's rankings of houses of the same type. Section~\ref{sec:cex} contains a complete example: Table~\ref{tab:Rstar} lists the strict preferences induced by a tie-breaking structure at several preference profiles, and the text following it runs the TTC algorithm at these profiles step by step. The table lists each agent's preferences only down to his own house; since an agent's own house remains available as long as he does, the houses he ranks below it play no role in the TTC algorithm.

Following \citet{SandholtzTai2026}, we do not require that an agent ranks himself first in his tie-breaking order; hence, an agent who prefers his own type may point to another agent's house of his own type.

To state the characterizations of \citet{Ma1994} and \citet{Ehlers2014} in our notation, we consider two further preference domains. For each coalition $N$, let $\Rd^s(H_N)$ denote the set of all linear orders over $H_N$ (\emph{strict preferences}) and let $\Rd^w(H_N)$ denote the set of all complete and transitive preferences over $H_N$ (\emph{weak preferences}). Problems, rules, and the properties introduced above are defined for these preferences in the same way, where, for weak preferences, two matchings for $N$ are welfare equivalent at a preference profile $R\in\Rd^w(H_N)^N$ if each agent $i\in N$ is indifferent according to $R_i$ between his allotments at the two matchings. For a coalition $N$, we say that $f(N,\cdot)$ satisfies a property if the requirements of the property hold at each problem with coalition $N$.

The \emph{TTC rule}, $\varphi^{T\!T\!C}$, assigns to each problem $(N,R)$ with $R\in\Rd^s(H_N)^N$ the outcome of the top trading cycles algorithm for $(N,R)$.

\begin{citedtheorem}[\citealp{Ma1994}]\label{thm:ma}
Let $N$ be a coalition and consider the domain of strict preferences, i.e., the set of preferences $\Rd^s(H_N)$. Then, for each rule $f$, $f(N,\cdot)$ satisfies \textsl{individual rationality}, \textsl{Pareto efficiency}, and \textsl{strategy-proofness} if and only if $f(N,\cdot)=\varphi^{T\!T\!C}(N,\cdot)$.
\end{citedtheorem}

If each type consists of a single house, then for each coalition $N$, $\Rd(H_N)=\Rd^s(H_N)$, and for each tie-breaking structure $\succ$, $\varphi^\succ=\varphi^{T\!T\!C}$. Hence, the model with strict preferences is the extreme case of our model in which each type consists of a single house.

Given two tie-breaking structures $\succ$ and $\widetilde\succ$, a problem $(N,R)$ with $R\in\Rd^w(H_N)^N$, and an agent $i\in N$, agent $i$'s strict preference $R^{\succ,\widetilde\succ}_i$ over $H_N$ ranks houses that are not indifferent according to $R_i$ as $R_i$ does, breaks ties within the indifference class of $R_i$ that contains $h(i)$ according to $\widetilde\succ_i$, and breaks ties within each other indifference class of $R_i$ according to $\succ_i$. The \emph{TTC rule based on $(\succ,\widetilde\succ)$}, $\varphi^{\succ,\widetilde\succ}$, assigns to each problem $(N,R)$ with $R\in\Rd^w(H_N)^N$ the outcome of the top trading cycles algorithm for $(N,R^{\succ,\widetilde\succ})$.

\begin{citedtheorem}[{\citealp[Theorem~2]{Ehlers2014}}]\label{thm:ehlers}
Consider the domain of weak preferences, i.e., for each coalition $N$, the set of preferences $\Rd^w(H_N)$.
\begin{enumerate}[label=\textup{(\alph*)},nosep]
\item For all tie-breaking structures $\succ$ and $\widetilde\succ$, $\varphi^{\succ,\widetilde\succ}$ satisfies \textsl{individual rationality}, \textsl{weak Pareto efficiency}, \textsl{strategy-proofness}, \textsl{non-bossiness}, and \textsl{consistency}.
\item If a rule $f$ satisfies \textsl{individual rationality}, \textsl{weak Pareto efficiency}, \textsl{strategy-proofness}, \textsl{non-bossiness}, and \textsl{consistency}, then there exist tie-breaking structures $\succ$ and $\widetilde\succ$ such that $f$ is welfare equivalent to $\varphi^{\succ,\widetilde\succ}$.
\end{enumerate}
\end{citedtheorem}

On the domain of objective indifferences, the indifference classes of agent $i$'s preferences are the sets of houses of the same type, and the indifference class that contains $h(i)$ is the set of houses of type $\tau(h(i))$. Hence, for all tie-breaking structures $\succ$ and $\widetilde\succ$, $\varphi^{\succ,\widetilde\succ}$ coincides on the domain of objective indifferences with $\varphi^{\succ^*}$, where, for each agent $i$, $\succ^*_i$ ranks the owners of houses of type $\tau(h(i))$ according to $\widetilde\succ_i$ and the owners of houses of each other type according to $\succ_i$; conversely, $\varphi^\succ$ coincides on this domain with $\varphi^{\succ,\succ}$. Thus, on the domain of objective indifferences, the class of rules characterized by Theorem~\ref{thm:ehlers} coincides with the class of TTC rules with fixed tie-breaking defined above. Theorem~\ref{thm:ehlers}(b), however, concerns rules defined for all weak preferences, and \citet[Theorem~3]{Ehlers2014} extends it to domains that contain all strict preferences. Except in the extreme case in which each type consists of a single house, the domain of objective indifferences does not contain all strict preferences. Hence, neither Theorem~\ref{thm:ehlers}(b) nor its extension in \citet[Theorem~3]{Ehlers2014} applies to the domain of objective indifferences.

For general weak preferences, TTC rules with fixed tie-breaking need not be \textsl{Pareto efficient} \citep{JaramilloManjunath2012,Ehlers2014} or \textsl{group strategy-proof} \citep{SandholtzTai2026}. On the domain of objective indifferences, \citet{SandholtzTai2026} show that TTC rules with fixed tie-breaking are \textsl{Pareto efficient} and \textsl{group strategy-proof}. We next record that, on this domain, TTC rules with fixed tie-breaking satisfy all properties introduced above.

\begin{lemma}[Properties of TTC rules with fixed tie-breaking]\label{lem:ttc}
Each TTC rule with fixed tie-breaking satisfies \textsl{individual rationality}, \textsl{Pareto efficiency}, \textsl{group strategy-proofness}, \textsl{non-bossiness}, and \textsl{consistency}.
\end{lemma}

\begin{proof}
Let $\succ$ be a tie-breaking structure. Since \citet[Theorems~1 and~2]{SandholtzTai2026} establish \textsl{Pareto efficiency} and \textsl{group strategy-proofness} for every tie-breaking structure, $\varphi^\succ$ satisfies \textsl{Pareto efficiency} and \textsl{group strategy-proofness}. The domain of objective indifferences is closed under restrictions to coalitions and is a subdomain of the domain of weak preferences. Let $\varphi^{\succ,\succ}$ denote the TTC rule for weak preferences based on $(\succ,\widetilde\succ)$ with $\widetilde\succ=\succ$, i.e., the rule that breaks all ties in the agents' preferences according to $\succ$. As noted after Theorem~\ref{thm:ehlers}, $\varphi^{\succ,\succ}$ coincides with $\varphi^\succ$ on the domain of objective indifferences. Hence, Theorem~\ref{thm:ehlers}(a) implies that $\varphi^\succ$ satisfies \textsl{individual rationality}, \textsl{non-bossiness}, and \textsl{consistency}.
\end{proof}

\section{A counterexample to extending the Ma and Ehlers characterizations}\label{sec:cex}

Throughout this section, let $\Nc=\{1,2,3,4,5\}$ and $\T=\{a,b\}$, where agents $1$, $2$, and $3$ own houses of type $a$ and agents $4$ and $5$ own houses of type $b$, i.e., the population has type composition $aaabb$. Since there are two types, for each coalition $N$ that contains owners of houses of both types, each agent $i\in N$ either prefers his own type to the other type or the other type to his own type.

Let $\succ$ be a tie-breaking structure in which each agent ranks himself first and such that
\[ 4\succ_1 5,\quad 4\succ_2 5,\quad 5\succ_3 4,\quad 1\succ_4 3\succ_4 2,\quad\text{and}\quad 1\succ_5 2\succ_5 3,\]
i.e., at each induced strict profile $R^\succ$, agents $1$ and $2$ rank $h(4)$ above $h(5)$, agent $3$ ranks $h(5)$ above $h(4)$, agent $4$ ranks the houses of type $a$ as $h(1),h(3),h(2)$, and agent $5$ ranks them as $h(1),h(2),h(3)$. Let $R^*\in\Rd(H_\Nc)^\Nc$ be the preference profile at which each agent prefers the other type to his own type, and, for each $i\in\{1,4,5\}$, let $R^i$ be the preference profile that differs from $R^*$ only in that agent $i$ prefers his own type.

\begin{table}[h!t]
\centering
\begin{tabular}{ccccc|c|c|c}
\multicolumn{5}{c|}{$R^*$} & $R^1$ & $R^4$ & $R^5$\\
agent $1$ & agent $2$ & agent $3$ & agent $4$ & agent $5$ & agent $1$ & agent $4$ & agent $5$\\
\hline
$h(4)$ & $\boxed{h(4)}$ & $\boxed{h(5)}$ & $h(1)$ & $h(1)$ & $h(1)$ & $h(4)$ & $h(5)$\\
$h(5)$ & $h(5)$ & $h(4)$ & $\boxed{h(3)}$ & $\boxed{h(2)}$ & & & \\
$\boxed{h(1)}$ & $h(2)$ & $h(3)$ & $h(2)$ & $h(3)$ & & & \\
       &        &        & $h(4)$ & $h(5)$ & & & \\
\end{tabular}
\caption{Strict preferences induced by $\succ$ at $R^*$, and those of agents $1$, $4$, and $5$ at $R^1$, $R^4$, and $R^5$, respectively, listed down to each agent's own house. At $R^i$, $i\in\{1,4,5\}$, all agents other than agent $i$ have the same preferences as at $R^*$. At $R^*$, the houses that the agents receive at the matching $\mu^*$ defined below are boxed.}\label{tab:Rstar}
\end{table}

Let $\mu^*$ be the matching for $\Nc$ with
\[ \mu^*(1)=h(1),\quad \mu^*(2)=h(4),\quad \mu^*(3)=h(5),\quad \mu^*(4)=h(3),\quad \mu^*(5)=h(2).\]
Define the rule $\psi$ by
\[ \psi(N,R)=\begin{cases}\mu^* & \text{if } (N,R)=(\Nc,R^*),\\ \varphi^{\succ}(N,R) & \text{otherwise.}\end{cases}\]
Thus, $\psi$ coincides with the TTC rule $\varphi^\succ$ at all problems except $(\Nc,R^*)$, where it selects $\mu^*$ instead of $\varphi^\succ(\Nc,R^*)$. Given the strict preferences listed in Table~\ref{tab:Rstar}, at Step~1 of the TTC algorithm for $\succ$ at $(\Nc,R^*)$, agent $1$ points to $h(4)$ and agent $4$ points to $h(1)$; hence, agents $1$ and $4$ trade. At Step~2, agent $5$ points to $h(2)$, his highest-ranked remaining house, and agents $2$ and $3$ point to $h(5)$, the only remaining house of type $b$; hence, agents $2$ and $5$ trade, and agent $3$ keeps his house. Thus, the trading cycles of $\varphi^{\succ}(\Nc,R^*)$ are
\[ 1\to h(4)\to 4\to h(1)\to 1,\quad 2\to h(5)\to 5\to h(2)\to 2,\quad\text{and}\quad 3\to h(3)\to 3,\]
whereas the trading cycles of $\psi(\Nc,R^*)=\mu^*$ are
\[
1\to h(1)\to 1\quad\text{and}\quad 2\to h(4)\to 4\to h(3)\to 3\to h(5)\to 5\to h(2)\to 2.
\]
The associated type allocations are
\[
\tau(\varphi^{\succ}(\Nc,R^*))=(b,b,a,a,a)\quad\text{and}\quad\tau(\mu^*)=(a,b,b,a,a);
\]
hence, agents $1$ and $3$ receive houses of different types at the two matchings, and the two matchings are not welfare equivalent.

\begin{lemma}[TTC outcomes at $R^1$ and at $R^*|_{\{2,3,4,5\}}$]\label{lem:comp}\leavevmode
\begin{enumerate}[label=\textup{(\alph*)},nosep]
\item $\varphi^{\succ}(\Nc,R^1)=\mu^*$.
\item $\varphi^{\succ}(\{2,3,4,5\},R^*|_{\{2,3,4,5\}})$ coincides with $\mu^*$ on $\{2,3,4,5\}$.
\end{enumerate}
\end{lemma}

\begin{proof}(a) Given the strict preferences listed in Table~\ref{tab:Rstar}, at Step~1 of the TTC algorithm for $\succ$ at $(\Nc,R^1)$, agents $1$, $4$, and $5$ point to $h(1)$, agent $2$ points to $h(4)$, and agent $3$ points to $h(5)$; hence, the only cycle is $1\to h(1)\to 1$. At Step~2, agent $4$ points to $h(3)$ and agent $5$ points to $h(2)$, while agents $2$ and $3$ still point to $h(4)$ and $h(5)$, respectively, which yields the cycle $2\to h(4)\to 4\to h(3)\to 3\to h(5)\to 5\to h(2)\to 2$. Hence, $\varphi^{\succ}(\Nc,R^1)=\mu^*$.\smallskip

\noindent (b) Without agent $1$ and $h(1)$, Step~1 of the TTC algorithm for $\succ$ at $(\{2,3,4,5\},R^*|_{\{2,3,4,5\}})$ coincides with Step~2 in (a), and the same cycle forms.
\end{proof}

\begin{theorem}[A five-agent counterexample]\label{thm:main}
The rule $\psi$ satisfies \textsl{individual rationality}, \textsl{Pareto efficiency}, \textsl{group strategy-proofness}, \textsl{non-bossiness}, and \textsl{consistency}. However, for each tie-breaking structure $\succ'$, $\psi(\Nc,\cdot)$ is not welfare equivalent to $\varphi^{\succ'}(\Nc,\cdot)$; in particular, $\psi$ is not welfare equivalent to any TTC rule with fixed tie-breaking.
\end{theorem}

\begin{proof}
By Lemma~\ref{lem:ttc}, $\varphi^{\succ}$ satisfies all five properties. Since $\psi$ differs from $\varphi^{\succ}$ only at $(\Nc,R^*)$, it suffices to check the requirements of the properties that involve the problem $(\Nc,R^*)$.\smallskip

\noindent\emph{Individual rationality and Pareto efficiency.} Since, at $\mu^*$, agent $1$ receives his own house and agents $2$, $3$, $4$, and $5$ receive houses of their preferred types, $\mu^*$ is \textsl{individually rational} for $R^*$. Since, at $\mu^*$, agents $2$, $3$, $4$, and $5$ receive houses of their preferred types, none of them can be made better off. Agent $1$ can only be made better off by receiving a house of type $b$. However, at $\mu^*$, both houses of type $b$ are assigned to agents $2$ and $3$. Hence, if agent $1$ receives one of them, then agent $2$ or agent $3$ receives a house of type $a$ instead, and this agent is worse off. Thus, $\mu^*$ is \textsl{Pareto efficient} for $R^*$.\smallskip

\noindent\emph{Group strategy-proofness.} Suppose, by contradiction, that there exist a problem $(N,R)$, a group $S\subseteq N$, and $R'_S\in\Rd(H_N)^S$ such that $S$ manipulates $\psi$ at $(N,R)$ via $R'_S$. Let $R'=(R'_S,R_{-S})$ and let $S'=\{i\in S:R'_i\neq R_i\}\subseteq S$ be the set of members of $S$ who change their preferences. Since some member of $S$ strictly gains, $R'\neq R$, i.e., $S'\neq\emptyset$. Since $\psi$ coincides with $\varphi^{\succ}$ except at $(\Nc,R^*)$ and $\varphi^{\succ}$ satisfies \textsl{group strategy-proofness}, $N=\Nc$ and $R^*\in\{R,R'\}$, i.e., $\mu^*=\psi(\Nc,R^*)$ is the outcome either at the true preferences $R$ or at the reported preferences $R'$. Since there are two types and preferences are strict across types, each agent $i\in S'$ ranks the two types at $R'_i$ in the opposite order from $R_i$.

First, suppose that $R=R^*$, i.e., $\psi(\Nc,R)=\mu^*$. Then, each agent $i\in S'$ prefers his own type at $R'_i$. Since $R'\neq R^*$, the definition of $\psi$ implies that $\psi(\Nc,R')=\varphi^{\succ}(\Nc,R')$, and since, by Lemma~\ref{lem:ttc}, $\varphi^{\succ}$ satisfies \textsl{individual rationality}, each agent $i\in S'$ receives a house of his own type at $\psi(\Nc,R')$. At $\mu^*$, each agent in $\Nc\setminus\{1\}$ receives a house of his preferred type at $R^*$, i.e., of the other type; hence, no agent in $S'\setminus\{1\}$ weakly gains. Since each member of $S$, and hence each member of $S'$, weakly gains, $S'\subseteq\{1\}$. Since $S'\neq\emptyset$, $S'=\{1\}$, i.e., $R'=R^1$. By Lemma~\ref{lem:comp}(a), $\psi(\Nc,R')=\varphi^{\succ}(\Nc,R^1)=\mu^*=\psi(\Nc,R)$; hence, no agent strictly gains, a contradiction.

Second, suppose that $R'=R^*$, i.e., $\psi(\Nc,R')=\mu^*$. Then, each agent $i\in S'$ truly prefers his own type at $R_i$. Since $R\neq R^*$, the definition of $\psi$ implies that $\psi(\Nc,R)=\varphi^\succ(\Nc,R)$, and since, by Lemma~\ref{lem:ttc}, $\varphi^\succ$ satisfies \textsl{individual rationality}, each agent $i\in S'$ receives a house of his own type at $\psi(\Nc,R)$. At $\mu^*$, each agent in $\Nc\setminus\{1\}$ receives a house of the other type; hence, no agent in $S'\setminus\{1\}$ weakly gains. Since each member of $S$, and hence each member of $S'$, weakly gains, $S'\subseteq\{1\}$. Since $S'\neq\emptyset$, $S'=\{1\}$, i.e., $R=R^1$. By Lemma~\ref{lem:comp}(a), $\psi(\Nc,R)=\varphi^{\succ}(\Nc,R^1)=\mu^*=\psi(\Nc,R')$; hence, no agent strictly gains, a contradiction.\smallskip

\noindent\emph{Non-bossiness.} Since $\psi$ coincides with $\varphi^{\succ}$ except at $(\Nc,R^*)$ and $\varphi^{\succ}$ satisfies \textsl{non-bossiness}, it suffices to compare $(\Nc,R^*)$ with the problems at which the preferences of a single agent differ from $R^*$. Let $i\in\Nc$, let $R'_i\in\Rd(H_\Nc)$ with $R'_i\neq R^*_i$, and let $R'=(R'_i,R^*_{-i})$. We check \textsl{non-bossiness} for both changes, from $R^*_i$ to $R'_i$ and from $R'_i$ to $R^*_i$. Since $R'\neq R^*$, the definition of $\psi$ implies that $\psi(\Nc,R')=\varphi^{\succ}(\Nc,R')$.

If $i=1$, then $R'=R^1$ and, by Lemma~\ref{lem:comp}(a), $\psi(\Nc,R')=\varphi^{\succ}(\Nc,R^1)=\mu^*=\psi(\Nc,R^*)$. Since the matchings at $R^*$ and at $R'$ coincide, \textsl{non-bossiness} is satisfied in both directions.

If $i\neq 1$, then agent $i$ prefers his own type at $R'_i$. Since $\varphi^{\succ}$ satisfies \textsl{individual rationality}, agent $i$ receives a house of his own type at $\psi(\Nc,R')$, whereas he receives a house of the other type at $\mu^*$. Hence, agent $i$'s allotment changes, and \textsl{non-bossiness} imposes no requirement in either direction. Thus, $\psi$ satisfies \textsl{non-bossiness}.\smallskip

\noindent\emph{Consistency.} Since a reduced problem has fewer agents than $\Nc$, the problem $(\Nc,R^*)$ appears in the requirements of \textsl{consistency} only as the original problem, never as a reduced problem. Since $\mu^*$ consists of the trading cycles $1\to h(1)\to 1$ and $2\to h(4)\to 4\to h(3)\to 3\to h(5)\to 5\to h(2)\to 2$, at $\mu^*$, the only non-empty groups $S\subsetneq\Nc$ whose allotments are precisely their endowments are $\{1\}$ and $\{2,3,4,5\}$. If $\{1\}$ leaves, then, since $\psi$ coincides with $\varphi^{\succ}$ at $(\{2,3,4,5\},R^*|_{\{2,3,4,5\}})$, Lemma~\ref{lem:comp}(b) implies that $\psi(\{2,3,4,5\},R^*|_{\{2,3,4,5\}})$ coincides with $\mu^*$ on $\{2,3,4,5\}$. If $\{2,3,4,5\}$ leaves, then agent $1$ receives his own house at $(\{1\},R^*|_{\{1\}})$. Hence, $\psi$ satisfies \textsl{consistency}.\smallskip

\noindent\emph{$\psi(\Nc,\cdot)$ is not welfare equivalent to a TTC rule with fixed tie-breaking.} Suppose, by contradiction, that for some tie-breaking structure $\succ'$, $\psi(\Nc,\cdot)$ is welfare equivalent to $\varphi^{\succ'}(\Nc,\cdot)$. Recall that, at $R^5$, agent $5$ prefers his own type and all other agents prefer the other type. Consider the TTC algorithm for $\succ'$ at $(\Nc,R^5)$, and let $x\in\{1,2,3\}$ be the agent whose house agent $4$ ranks highest among the houses of type $a$ under $\succ'_4$. Since agent $5$ prefers his own type and $\varphi^{\succ'}$ satisfies \textsl{individual rationality}, agent $5$ receives a house of type $b$. Hence, at most one agent of $\{1,2,3\}$ receives a house of type $b$.

Since agent $4$ prefers type $a$, he points to $h(x)$ as long as agent $x$ remains. Since $h(x)$ points to its owner $x$, every cycle containing agent $4$ while agent $x$ remains must also contain agent $x$. Hence, agent $4$ cannot be removed before agent $x$, and $h(4)$ remains as long as agent $x$ remains. Since agent $x$ prefers type $b$ and $h(4)$ remains as long as agent $x$ remains, agent $x$ points to a house of type $b$ at every step at which he remains. Therefore, when agent $x$ is removed in a cycle, he receives a house of type $b$. Since at most one agent of $\{1,2,3\}$ can receive a house of type $b$, at $\varphi^{\succ'}(\Nc,R^5)$, agent $x$ is the unique such agent.

We next identify agent $x$ by using that $\psi(\Nc,\cdot)$ and $\varphi^{\succ'}(\Nc,\cdot)$ are welfare equivalent. Since $R^5\neq R^*$, the definition of $\psi$ implies that $\psi(\Nc,R^5)=\varphi^{\succ}(\Nc,R^5)$. Given the strict preferences listed in Table~\ref{tab:Rstar}, at Step~1 of the TTC algorithm for $\succ$ at $(\Nc,R^5)$, agent $1$ points to $h(4)$ and agent $4$ points to $h(1)$. Hence, agents $1$ and $4$ trade at Step~1, and agent $1$ receives a house of type $b$ at $\psi(\Nc,R^5)$. By welfare equivalence, agent $1$ also receives a house of type $b$ at $\varphi^{\succ'}(\Nc,R^5)$. Since agent $x$ is the unique agent of $\{1,2,3\}$ who receives a house of type $b$ at $\varphi^{\succ'}(\Nc,R^5)$, $x=1$. Thus, agent $4$ ranks $h(1)$ highest among the houses of type $a$ under $\succ'_4$.

Similarly, recall that, at $R^4$, agent $4$ prefers his own type and all other agents prefer the other type. By the same argument with the roles of agents $4$ and $5$ exchanged, the unique agent of $\{1,2,3\}$ who receives a house of type $b$ at $\varphi^{\succ'}(\Nc,R^4)$ is the agent whose house agent $5$ ranks highest among the houses of type $a$ under $\succ'_5$. Since $R^4\neq R^*$, the definition of $\psi$ implies that $\psi(\Nc,R^4)=\varphi^{\succ}(\Nc,R^4)$. Given the strict preferences listed in Table~\ref{tab:Rstar}, at Step~1 of the TTC algorithm for $\succ$ at $(\Nc,R^4)$, agents $1$, $2$, and $4$ point to $h(4)$, agent $3$ points to $h(5)$, and agent $5$ points to $h(1)$. Hence, the only cycle at Step~1 is $4\to h(4)\to 4$. At Step~2, agent $1$ points to $h(5)$, the only remaining house of type $b$, and agent $5$ points to $h(1)$, his highest-ranked house of type $a$. Hence, agents $1$ and $5$ trade at Step~2, and agent $1$ receives a house of type $b$ at $\psi(\Nc,R^4)$. As before, it follows that agent $5$ ranks $h(1)$ highest among the houses of type $a$ under $\succ'_5$.

Now consider the TTC algorithm for $\succ'$ at $(\Nc,R^*)$. Since agents $4$ and $5$ prefer type $a$ and both rank $h(1)$ highest among the houses of type $a$, they point to $h(1)$ at Step~1. Since agent $1$ prefers type $b$, he points to $h(4)$ or $h(5)$. Hence, agent $1$ trades with agent $4$ or agent $5$ at Step~1 and receives a house of type $b$ at $\varphi^{\succ'}(\Nc,R^*)$. However, agent $1$ receives a house of type $a$ at $\psi(\Nc,R^*)=\mu^*$, a contradiction.
\end{proof}

Hence, by Theorem~\ref{thm:main}, neither the characterization in Theorem~\ref{thm:ma} nor the one in Theorem~\ref{thm:ehlers} extends to the domain of objective indifferences. Relative to \citet{Ehlers2014}, the counterexample satisfies the stronger requirements of \textsl{Pareto efficiency} and \textsl{group strategy-proofness} in place of \textsl{weak Pareto efficiency} and \textsl{strategy-proofness}, while retaining \textsl{individual rationality}, \textsl{non-bossiness}, and \textsl{consistency}. Relative to \citet{Ma1994}, the fixed-population rule $\psi(\Nc,\cdot)$ satisfies \textsl{individual rationality}, \textsl{Pareto efficiency}, and the stronger property of \textsl{group strategy-proofness}, yet is not welfare equivalent to any TTC rule with fixed tie-breaking.

\begin{remark}[Minimality for a fixed population]\label{rem:min}
With two types of houses, five agents are the smallest number of agents for which a fixed-population counterexample as in Theorem~\ref{thm:main} exists. Up to relabeling of the types, the two type compositions of five agents using both types are $aaaab$ and $aaabb$. Theorem~\ref{thm:main} provides a counterexample for a population with type composition $aaabb$. For type composition $aaaab$ and for each two-type population with at most four agents, the computer verification in Appendix~\ref{app:search} shows that no fixed-population counterexample exists. That is, for each of these populations $\Nc$ and each rule $f$ that satisfies \textsl{individual rationality}, \textsl{Pareto efficiency}, and \textsl{group strategy-proofness}, there is a tie-breaking structure $\succ$ such that $f(\Nc,\cdot)$ is welfare equivalent to $\varphi^\succ(\Nc,\cdot)$. Thus, five agents are minimal for the fixed-population result. Since with one type all matchings are welfare equivalent, two types are also minimal. The computer verification concerns only the problems with all agents of the population; a four-agent counterexample that relies on the behavior of the rule on a smaller coalition does exist, and we discuss it next.

For the characterization in Theorem~\ref{thm:ehlers}, a weaker counterexample suffices. There, a rule is defined for all coalitions of potential agents, and welfare equivalence to a TTC rule with fixed tie-breaking requires a single tie-breaking structure that works for all coalitions simultaneously. Hence, a rule may fail to be welfare equivalent to any TTC rule with fixed tie-breaking even though, for each coalition, its restriction to that coalition is welfare equivalent to the restriction of some TTC rule with fixed tie-breaking. In Appendix~\ref{app:four}, we construct, for a population with type composition $aabb$, such a rule that satisfies the properties in Theorem~\ref{thm:main}: its restrictions to the coalition of all four agents and to the coalition $\{1,2,3\}$ are welfare equivalent to restrictions of TTC rules with fixed tie-breaking, but these two coalitions require incompatible tie-breaking orders for agent $3$. With two types, four potential agents are minimal for such a counterexample (Remark~\ref{rem:min4}).
\end{remark}

\section{Robustness and conclusion}\label{sec:robust}

In this section, we show that the rule $\psi$ of Section~\ref{sec:cex} is also a counterexample for several other characterizations of TTC for strict preferences.

In addition to \citet{Ma1994}, several authors characterize the TTC rule on the domain of strict preferences by other properties. Since the model with strict preferences is the extreme case of our model in which each type consists of a single house, the same question arises for each of these characterizations as for Theorem~\ref{thm:ma}: do the corresponding properties characterize TTC rules with fixed tie-breaking, up to welfare equivalence, on the domain of objective indifferences? Most of these properties compare allotments only through the agents' preferences and therefore apply to our domain without change. Other properties refer to particular houses or exact allotments; for these we specify the adaptations considered below, generally in terms of types. We show that $\psi$ satisfies the corresponding requirements. Since $\psi(\Nc,\cdot)$ is not welfare equivalent to any TTC rule with fixed tie-breaking (Theorem~\ref{thm:main}), the characterizations considered here do not extend under these formulations.
\begin{enumerate}[label=(\roman*),wide,labelindent=0pt,listparindent=\parindent,topsep=\medskipamount,itemsep=\medskipamount,parsep=0pt]
\item \emph{Pair-efficiency} requires that no two agents can both strictly gain by exchanging their allotments. Since \textsl{Pareto efficiency} implies \textsl{pair-efficiency}, the characterization of the TTC rule by \textsl{individual rationality}, \textsl{pair-efficiency}, and \textsl{strategy-proofness} \citep{Ekici2024,EkiciSethuraman2024} does not extend. With two types, as in our counterexample, the two properties are equivalent: in any Pareto improvement, some agent moves from a house of his less preferred type to a house of his preferred type, and some other agent moves from a house of the latter type to a house of the other type; since he is not worse off, he strictly prefers the other type, and these two agents can exchange their allotments and both strictly gain.
\item \emph{Drop strategy-proofness}, \emph{truncation-proofness}, and \emph{upper invariance} weaken \textsl{strategy-proofness}. The first two properties require only that no agent can gain by a particular kind of misreport: a drop or a truncation, respectively. Under strict preferences, a drop moves one house other than the endowment to the bottom of the ranking, whereas a truncation places the endowment above a lower segment of the other houses, preserving their relative order. Thus, the endowment moves up relative to any houses in that segment that were previously ranked above it. On our domain, we apply these deviations to types rather than to individual houses: a drop moves one type other than the agent's own type to the bottom, and a truncation places his endowment type above a lower segment of the other types, again preserving their relative order. \textsl{Upper invariance} requires an agent's allotment to remain unchanged when he reorders only houses ranked below it; here we require the type of his allotment to remain unchanged when he reorders only lower-ranked types. All three adapted properties follow from \textsl{strategy-proofness}. Hence, the following characterizations of the TTC rule do not extend: by \textsl{individual rationality}, \textsl{Pareto efficiency}, and \textsl{drop strategy-proofness} or \textsl{upper invariance} \citep{AltuntasPhanTamura2023}; by \textsl{individual rationality}, \textsl{Pareto efficiency}, and \textsl{truncation-proofness} \citep{CorenoFeng2026}; and by \textsl{individual rationality}, \textsl{pair-efficiency}, and \textsl{upper invariance} \citep{ChenJiaoZhangZhang2025}.
\item \emph{Ontoness} requires that each matching is selected at some preference profile. On the domain of objective indifferences, this requirement is too demanding; for instance, for a population with five agents and two types, there are $2^5=32$ preference profiles but $5!=120$ matchings. The natural requirement is \emph{type ontoness}: each type allocation that is induced by some matching is selected at some preference profile. At the preference profile at which each agent ranks first the type that a given type allocation assigns to him, \textsl{Pareto efficiency} requires that each agent receives that type; hence, \textsl{Pareto efficiency} implies \textsl{type ontoness}. Thus, with \textsl{ontoness} formulated as \textsl{type ontoness}, the characterizations of the TTC rule by \textsl{individual rationality}, \textsl{group strategy-proofness}, and \textsl{ontoness}, and by \textsl{individual rationality}, \textsl{strategy-proofness}, \textsl{non-bossiness}, and \textsl{ontoness} \citep{Takamiya2001} do not extend.
\item \citet[Lemma~1]{Ehlers2014} characterizes the TTC rule for strict preferences by \textsl{individual rationality}, \textsl{weak Pareto efficiency}, \textsl{strategy-proofness}, and \textsl{consistency}. Unlike the characterization in Theorem~\ref{thm:ehlers}, this characterization of the TTC rule does not use \textsl{non-bossiness}. Since $\psi$ satisfies these properties, this characterization does not extend either.
\item \citet{FujinakaWakayama2018} characterize the TTC rule for strict preferences by \textsl{individual rationality}, \textsl{strategy-proofness}, and \textsl{endowments-swapping-proofness}, which requires that no two agents can both strictly gain by swapping their endowments before the rule is applied. \textsl{Endowments-swapping-proofness} is also used in the characterizations of the TTC rule by \textsl{individual rationality}, \textsl{endowments-swapping-proofness}, and \textsl{upper invariance} \citep{ChenJiaoZhangZhang2025} and by \textsl{individual rationality}, \textsl{endowments-swapping-proofness}, and \textsl{truncation-proofness} \citep{CorenoFeng2025}. Since this property compares problems with different assignments of houses to agents, we extend the model for the following analysis. As in Section~\ref{sec:cex}, let $\Nc=\{1,2,3,4,5\}$ and $\T=\{a,b\}$, where houses $h(1)$, $h(2)$, and $h(3)$ are of type $a$ and houses $h(4)$ and $h(5)$ are of type $b$. An \emph{endowment assignment} is a bijection $\omega$ from $\Nc$ to $H_\Nc$, where agent $i$ owns house $\omega(i)$; let $\omega^0$ denote the original endowment assignment, i.e., for each $i\in\Nc$, $\omega^0(i)=h(i)$. A problem is now a pair $(R,\omega)$ of a preference profile $R\in\Rd(H_\Nc)^\Nc$ and an endowment assignment $\omega$, and a rule assigns a matching for $\Nc$ to each problem. For each endowment assignment $\omega$ and each pair of distinct agents $i,j\in\Nc$, let $\omega^{ij}$ be obtained from $\omega$ by swapping $\omega(i)$ and $\omega(j)$. A rule $f$ satisfies \emph{endowments-swapping-proofness} if there is no problem $(R,\omega)$ and no pair of distinct agents $i,j\in\Nc$ such that $f_i(R,\omega^{ij})\,P_i\,f_i(R,\omega)$ and $f_j(R,\omega^{ij})\,P_j\,f_j(R,\omega)$.

To define TTC rules with fixed tie-breaking in this extended model, we attach the tie-breaking orders to houses: given a tie-breaking structure $\succ$, each agent $i$ breaks ties between houses $h(k)$ and $h(l)$ of the same type according to $\succ_i$, independently of who owns these houses. For each problem $(R,\omega)$, let $\varphi^\succ(R,\omega)$ denote the outcome of the TTC algorithm for $R^\succ$ in which each house points to its owner under $\omega$. Since $R^\succ$ refines $R$, an agent who strictly gains according to $R$ also strictly gains according to $R^\succ$. Hence, since the TTC rule satisfies \textsl{endowments-swapping-proofness} for strict preferences \citep{FujinakaWakayama2018}, $\varphi^\succ$ satisfies \textsl{endowments-swapping-proofness}.

Recall from Section~\ref{sec:cex} that the rule $\psi$ coincides with the TTC rule $\varphi^\succ$ except at the problem $(\Nc,R^*)$, at which each agent $i$ prefers the other type to the type of $h(i)$; there, $\psi$ selects the matching $\mu^*$, at which agent $1$ keeps his house and agents $2$, $3$, $4$, and $5$ receive houses of their preferred types. We extend the rule $\psi$ by setting $\psi(R,\omega)=\mu^*$ if $(R,\omega)=(R^*,\omega^0)$ and $\psi(R,\omega)=\varphi^\succ(R,\omega)$ otherwise.

\begin{proposition}[Endowments-swapping-proofness]\label{prop:esp}
The extended rule $\psi$ satisfies \textsl{endowments-swapping-proofness}.
\end{proposition}

\begin{proof}
Since $\psi$ coincides with $\varphi^\succ$ except at $(R^*,\omega^0)$, it suffices to consider swaps from and to $(R^*,\omega^0)$. First, consider a swap from $(R^*,\omega^0)$, i.e., at the problem $(R^*,\omega^0)$, two distinct agents $i,j\in\Nc$ swap their endowments, which yields the problem $(R^*,(\omega^0)^{ij})$. Since, at $\mu^*$, each agent in $\{2,3,4,5\}$ receives a house of his preferred type and each pair of agents contains such an agent, no two agents both strictly gain.

Second, consider a swap to $(R^*,\omega^0)$, i.e., let $i,j\in\Nc$ be distinct and consider the problem $(R^*,\omega)$ with $\omega=(\omega^0)^{ij}$, at which agents $i$ and $j$, by swapping their endowments, obtain the problem $(R^*,\omega^0)$. If $1\in\{i,j\}$, then, at $\mu^*$, agent $1$ receives a house of type $a$, his less preferred type at $R^*$, and hence he cannot strictly gain. Now suppose that $1\notin\{i,j\}$. If $h(i)$ and $h(j)$ have different types, then, at $(R^*,\omega)$, both agents own houses of their preferred types. Since $\varphi^\succ$ satisfies \textsl{individual rationality}, at $\varphi^\succ(R^*,\omega)$, both receive houses of their preferred types, and therefore neither strictly gains at $\mu^*$.

It remains to consider the case in which $h(i)$ and $h(j)$ have the same type $t$. Since $1\notin\{i,j\}$, at $\mu^*$, both agents receive houses of their preferred type. Thus, at $\varphi^\succ(R^*,\omega)$, they must both receive houses of type $t$, their less preferred type, in order for both to strictly gain when the endowments are swapped back to $\omega^0$. Each original owner of a house of the other type prefers type $t$ at $R^*$. If any such owner did not receive a house of type $t$ at $\varphi^\succ(R^*,\omega)$, he could exchange his allotment with agent $i$ or with agent $j$, each of whom receives a house of type $t$, and both agents in this exchange would strictly gain. By \textsl{Pareto efficiency}, every original owner of a house of the other type must therefore receive a house of type $t$. Hence, the number of houses of type $t$ would have to be at least the number of houses of the other type plus two. Since there are three houses of type $a$ and two houses of type $b$, this is impossible.
\end{proof}

For each endowment assignment $\omega\neq\omega^0$, $\psi(\cdot,\omega)$ is a TTC rule with fixed tie-breaking, and $\psi(\cdot,\omega^0)$ is the rule $\psi(\Nc,\cdot)$. Hence, by Lemma~\ref{lem:ttc} and the arguments in Section~\ref{sec:cex}, for each endowment assignment, $\psi$ satisfies \textsl{individual rationality}, \textsl{strategy-proofness}, \textsl{upper invariance}, and \textsl{truncation-proofness}. Since, by Theorem~\ref{thm:main}, $\psi(\cdot,\omega^0)$ is not welfare equivalent to any TTC rule with fixed tie-breaking, none of the three characterizations of the TTC rule that use \textsl{endowments-swapping-proofness} extends to a characterization of TTC rules with fixed tie-breaking.
\end{enumerate}

\subsection*{Concluding remarks}

On the domain of objective indifferences, TTC rules with fixed tie-breaking regain \textsl{Pareto efficiency} and \textsl{group strategy-proofness} \citep{SandholtzTai2026}. Nevertheless, neither the properties that characterize the TTC rule for strict preferences \citep{Ma1994} nor those that characterize TTC rules with fixed tie-breaking for weak preferences \citep{Ehlers2014} single out TTC rules with fixed tie-breaking on this domain: the rule $\psi$ satisfies \textsl{individual rationality}, \textsl{Pareto efficiency}, \textsl{group strategy-proofness}, \textsl{non-bossiness}, and \textsl{consistency}, but already for a fixed population of five agents it is not welfare equivalent to any TTC rule with fixed tie-breaking. The same rule shows that several other characterizations of TTC for strict preferences do not extend either, including those based on \textsl{endowments-swapping-proofness}.

We do not consider all existing characterizations of TTC for strict preferences here. In particular, we leave aside the characterizations by \citet{ChenZhao2021}, \citet{KlausKlijnSethuraman2025}, and \citet{HuZhang2025}, whose properties (\textsl{rank monotonicity}, \textsl{respecting-improvement}, and \textsl{local unanimity}, respectively) require a more substantive adaptation when houses of the same type are identical. A characterization of TTC rules with fixed tie-breaking on the domain of objective indifferences, which \citet{SandholtzTai2026} leave for future work, requires properties beyond those used in the existing characterizations that we consider.

Our formulation of \textsl{consistency} is that of \citet{Ehlers2014}: a group can leave only with its own endowments. Since houses of the same type are identical, one can also allow a group to leave with other houses, as long as each remaining agent can still be endowed with a house of his own type, and require only that each remaining agent receive a house of the same type as before. \citet{KarakayaKlaus2026} study this type-based formulation of \textsl{consistency}. With two types of houses, a TTC rule with fixed tie-breaking in which each agent ranks himself first satisfies type-based \textsl{consistency} if and only if any two agents who own houses of the same type break ties among the houses of the other type in the same way, except possibly for the order of the two houses they rank highest \citep[Corollary~1]{KarakayaKlaus2026}; hence, with two types, \textsl{individual rationality}, \textsl{Pareto efficiency}, \textsl{group strategy-proofness}, and type-based \textsl{consistency} are compatible. With three types of houses, two of which are each owned by at least two agents, no rule satisfies \textsl{individual rationality}, \textsl{strategy-proofness}, type-based \textsl{consistency}, and either \textsl{weak Pareto efficiency} or \textsl{pair-efficiency} \citep[Proposition~3]{KarakayaKlaus2026}.

\appendix
\makeatletter
\renewcommand{\@seccntformat}[1]{Appendix \csname the#1\endcsname:\quad}
\makeatother

\section{Group strategy-proofness and welfare non-bossiness}\label{app:gsp}

\begin{proof}[Proof of Proposition~\ref{prop:gsp}]
First, let $f$ be \textsl{group strategy-proof}. Since an agent $i$ manipulates $f$ at a problem $(N,R)$ via $R'_i$ if and only if the group $\{i\}$ manipulates $f$ at $(N,R)$ via $R'_i$, $f$ is \textsl{strategy-proof}. Let $(N,R)$ be a problem, let $i\in N$, let $R'_i\in\Rd(H_N)$, let $R'=(R'_i,R_{-i})$, and suppose that $f_i(N,R')\,I_i\,f_i(N,R)$. Since preferences are strict across types, $f_i(N,R')$ and $f_i(N,R)$ have the same type. Hence, agent $i$ is indifferent between these two allotments both according to $R_i$ and according to $R'_i$. Suppose, by contradiction, that for some $j\in N$, $f_j(N,R')$ and $f_j(N,R)$ are not indifferent according to $R_j$; then $j\neq i$. If $f_j(N,R')\,P_j\,f_j(N,R)$, then the group $\{i,j\}$ manipulates $f$ at $(N,R)$ via $(R'_i,R_j)$. If $f_j(N,R)\,P_j\,f_j(N,R')$, then the group $\{i,j\}$ manipulates $f$ at $(N,R')$ via $(R_i,R_j)$. Both cases contradict \textsl{group strategy-proofness}. Hence, $f$ is \textsl{welfare non-bossy}.

Second, let $f$ be \textsl{strategy-proof} and \textsl{welfare non-bossy}, and suppose, by contradiction, that a group $S=\{i_1,\ldots,i_k\}$ manipulates $f$ at a problem $(N,R)$ via $R'_S$; let $R'=(R'_S,R_{-S})$. Since each member of $S$ weakly gains and preferences are strict across types, for each $i\in S$, either $\tau(f_i(N,R'))=\tau(f_i(N,R))$ or $f_i(N,R')\,P_i\,f_i(N,R)$. We show that $f(N,R)$ and $f(N,R')$ are welfare equivalent. Then no member of $S$ strictly gains, a contradiction.

To this end, we let the members of $S$ switch, one at a time, to auxiliary preferences, both starting from $R$ and starting from $R'$, until both sequences reach the same preference profile. For each $i\in S$, let $R''_i\in\Rd(H_N)$ rank type $\tau(f_i(N,R'))$ first and, if $\tau(f_i(N,R))\neq\tau(f_i(N,R'))$, type $\tau(f_i(N,R))$ second, with all remaining types ranked arbitrarily below these types. For each $l\in\{0,\ldots,k\}$, let $R^l$ be obtained from $R$ by replacing $R_{i_1},\ldots,R_{i_l}$ with $R''_{i_1},\ldots,R''_{i_l}$, and let $\widehat R^l$ be obtained from $R'$ by replacing $R'_{i_1},\ldots,R'_{i_l}$ with $R''_{i_1},\ldots,R''_{i_l}$. Hence, $R^0=R$ and $\widehat R^0=R'$, and, since $R$ and $R'$ coincide outside $S$, $R^k=\widehat R^k$. For each $l\in\{1,\ldots,k\}$, the preference profiles $R^{l-1}$ and $R^l$, as well as $\widehat R^{l-1}$ and $\widehat R^l$, differ only in the preferences of agent $i_l$. Hence, by \textsl{welfare non-bossiness}, if agent $i_l$ receives houses of the same type before and after such a switch, then the switch does not change any agent's welfare.\smallskip

\noindent\emph{Switching from $R$.} We show by induction that for each $l\in\{0,\ldots,k\}$, $f(N,R^l)$ is welfare equivalent to $f(N,R)$. For $l=0$, this holds trivially. Let $l\in\{1,\ldots,k\}$ and suppose that $f(N,R^{l-1})$ is welfare equivalent to $f(N,R)$. Then, at $f(N,R^{l-1})$, agent $i_l$ receives a house of type $t=\tau(f_{i_l}(N,R))$. Suppose, by contradiction, that at $f(N,R^l)$, agent $i_l$ receives a house of a type different from $t$. Since $R''_{i_l}$ ranks only type $\tau(f_{i_l}(N,R'))$ possibly above type $t$, there are two cases. If agent $i_l$ receives a house of type $\tau(f_{i_l}(N,R'))\neq t$, then, since $f_{i_l}(N,R')\,P_{i_l}\,f_{i_l}(N,R)$, agent $i_l$ manipulates $f$ at $(N,R^{l-1})$, where his preferences are $R_{i_l}$, via $R''_{i_l}$. Otherwise, agent $i_l$ receives a house of a type that $R''_{i_l}$ ranks below type $t$, and he manipulates $f$ at $(N,R^l)$, where his preferences are $R''_{i_l}$, via $R_{i_l}$. Both cases contradict \textsl{strategy-proofness}. Hence, at $f(N,R^l)$, agent $i_l$ receives a house of type $t$, and, by \textsl{welfare non-bossiness}, $f(N,R^l)$ is welfare equivalent to $f(N,R^{l-1})$, and hence to $f(N,R)$.\smallskip

\noindent\emph{Switching from $R'$.} We show by induction that for each $l\in\{0,\ldots,k\}$, $f(N,\widehat R^l)$ is welfare equivalent to $f(N,R')$. For $l=0$, this holds trivially. Let $l\in\{1,\ldots,k\}$ and suppose that $f(N,\widehat R^{l-1})$ is welfare equivalent to $f(N,R')$. Then, at $f(N,\widehat R^{l-1})$, agent $i_l$ receives a house of type $t'=\tau(f_{i_l}(N,R'))$. Since $R''_{i_l}$ ranks type $t'$ first, if, at $f(N,\widehat R^l)$, agent $i_l$ received a house of a type different from $t'$, then he would manipulate $f$ at $(N,\widehat R^l)$, where his preferences are $R''_{i_l}$, via $R'_{i_l}$, which contradicts \textsl{strategy-proofness}. Hence, at $f(N,\widehat R^l)$, agent $i_l$ receives a house of type $t'$, and, by \textsl{welfare non-bossiness}, $f(N,\widehat R^l)$ is welfare equivalent to $f(N,\widehat R^{l-1})$, and hence to $f(N,R')$.\smallskip

\noindent\emph{Conclusion.} Since $R^k=\widehat R^k$, $f(N,R)$ is welfare equivalent to $f(N,R^k)=f(N,\widehat R^k)$, which is welfare equivalent to $f(N,R')$. Hence, no member of $S$ strictly gains, a contradiction.

Third, we show that \textsl{group strategy-proofness} does not imply \textsl{non-bossiness} and that \textsl{strategy-proofness} together with \textsl{non-bossiness} does not imply \textsl{group strategy-proofness}. Let $\Nc=\{1,2,3\}$ and $\T=\{a,b\}$, where agents $1$ and $2$ own houses of type $a$ and agent $3$ owns a house of type $b$. In both rules below, each agent keeps his own house at each problem with fewer than three agents. Since there are two types, each agent's preferences are determined by his preferred type; for each preference profile $R$ for $\Nc$, we therefore write $R=(t_1,t_2,t_3)$, where $t_i$ is agent $i$'s preferred type at $R_i$.

Let $\varphi$ be the rule under which each agent keeps his own house, except that, whenever agent $3$ prefers type $a$, agents $1$ and $2$ exchange their houses. Since each agent receives a house of his own type at each problem, no agent's welfare depends on the reported preferences; hence, $\varphi$ is \textsl{group strategy-proof}. Since, at $(a,a,a)$ and $(a,a,b)$, agent $3$ receives $h(3)$ but agents $1$ and $2$ receive different houses, $\varphi$ is not \textsl{non-bossy}.

Let $\varphi'$ be the rule under which each agent keeps his own house, except at the following preference profiles: at $(b,a,a)$, agents $1$ and $3$ exchange their houses; at $(b,b,a)$, agent $1$ receives $h(2)$, agent $2$ receives $h(3)$, and agent $3$ receives $h(1)$. Since, at each problem, each agent receives a house of his own type or of his preferred type, $\varphi'$ is \textsl{individually rational}. Agent $3$'s report affects the matching only if $t_1=b$; in this case, if he reports type $a$ as preferred, then he receives a house of type $a$, and otherwise he receives his own house. Agent $2$'s report affects the matching only if $(t_1,t_3)=(b,a)$; in this case, if he reports type $a$ as preferred, then he receives a house of type $a$, and otherwise he receives a house of type $b$. Agent $1$'s report affects the matching only if $t_3=a$; in this case, if $t_2=a$ and he reports type $b$ as preferred, then he receives a house of type $b$; if $t_2=a$ and he reports type $a$ as preferred, then he receives his own house; and if $t_2=b$, then he receives a house of type $a$ under each report. Hence, no agent can gain by misreporting, and $\varphi'$ is \textsl{strategy-proof}. Moreover, whenever an agent's report affects the matching, it changes his own allotment; hence, $\varphi'$ is \textsl{non-bossy}. However, at $(a,b,a)$, each agent keeps his own house, and if agent $1$ reports type $b$ as preferred, then agent $1$ receives $h(2)$, a house of his preferred type $a$, and agent $2$ receives $h(3)$, a house of his preferred type $b$. Hence, the group $\{1,2\}$ manipulates $\varphi'$ at $(a,b,a)$ via agent $1$'s report and agent $2$'s true preferences, and $\varphi'$ is not \textsl{group strategy-proof}.
\end{proof}

\section{A four-agent counterexample}\label{app:four}

Throughout this appendix, let $\Nc=\{1,2,3,4\}$ and $\T=\{a,b\}$, where agents $1$ and $2$ own houses of type $a$ and agents $3$ and $4$ own houses of type $b$. Since there are two types, for each coalition $N$ that contains owners of houses of both types, each agent $i\in N$ either prefers his own type to the other type or the other type to his own type.

Let $\succ$ be a tie-breaking structure in which each agent ranks himself first and such that
\[
3\succ_1 4,\quad 3\succ_2 4,\quad 1\succ_3 2,\quad\text{and}\quad 1\succ_4 2,
\]
i.e., at each induced strict profile $R^\succ$, agents $1$ and $2$ rank $h(3)$ above $h(4)$, and agents $3$ and $4$ rank $h(1)$ above $h(2)$.

Let $\widetilde R\in\Rd(H_\Nc)^\Nc$ be the preference profile at which agents $1$, $2$, and $3$ prefer the other type to their own type and agent $4$ prefers his own type to the other type. Let $\widetilde R^1$, $\widetilde R^4$, and $\widetilde R^{14}$ be the preference profiles that differ from $\widetilde R$ only in that agent $1$ prefers type $a$ to type $b$, only in that agent $4$ prefers type $a$ to type $b$, and only in that agents $1$ and $4$ both prefer type $a$ to type $b$, respectively.

\begin{table}[h!t]
\centering
\begin{tabular}{cccc|c|c|cc}
\multicolumn{4}{c|}{$\widetilde R$} & $\widetilde R^1$ & $\widetilde R^4$ & \multicolumn{2}{c}{$\widetilde R^{14}$}\\
agent $1$ & agent $2$ & agent $3$ & agent $4$ & agent $1$ & agent $4$ & agent $1$ & agent $4$\\
\hline
$h(3)$ & $h(3)$ & $\boxed{h(1)}$ & $h(4)$ & $h(1)$ & $h(1)$ & $h(1)$ & $h(1)$\\
$h(4)$ & $\boxed{h(4)}$ & $h(2)$ & $\boxed{h(3)}$ &        & $h(2)$ &        & $h(2)$\\
$h(1)$ & $h(2)$ & $h(3)$ &        &        & $h(4)$ &        & $h(4)$\\
$\boxed{h(2)}$ &        &        &        &        &        &        &       \\
\end{tabular}
\caption{Strict preferences induced by $\succ$ at $\widetilde R$, and those of agent $1$ at $\widetilde R^1$, of agent $4$ at $\widetilde R^4$, and of agents $1$ and $4$ at $\widetilde R^{14}$, listed down to each agent's own house or, at $\widetilde R$, down to the house that the agent receives at the matching $\widetilde\mu$ defined below, whichever he ranks lower. At $\widetilde R^1$, $\widetilde R^4$, and $\widetilde R^{14}$, all other agents have the same preferences as at $\widetilde R$. At $\widetilde R$, the houses that the agents receive at $\widetilde\mu$ are boxed; agents $1$ and $4$ are indifferent between these houses and their own houses.}\label{tab:Rtilde}
\end{table}

Let $\widetilde\mu$ be the matching for $\Nc$ with
\[
\widetilde\mu(1)=h(2),\quad \widetilde\mu(2)=h(4),\quad \widetilde\mu(3)=h(1),\quad \widetilde\mu(4)=h(3).
\]
At $\widetilde\mu$, agents $1$ and $4$ receive houses of their own type but not their own houses; if an agent does not rank himself first in his tie-breaking order, TTC rules with fixed tie-breaking may also assign such allotments. Define the rule $\widetilde\psi$ by
\[
\widetilde\psi(N,R)=\begin{cases}\widetilde\mu & \text{if } (N,R)=(\Nc,\widetilde R),\\ \varphi^\succ(N,R) & \text{otherwise.}\end{cases}
\]
Thus, $\widetilde\psi$ coincides with the TTC rule $\varphi^\succ$ at all problems except $(\Nc,\widetilde R)$, where it selects $\widetilde\mu$ instead of $\varphi^\succ(\Nc,\widetilde R)$. Given the strict preferences listed in Table~\ref{tab:Rtilde}, at Step~1 of the TTC algorithm for $\succ$ at $(\Nc,\widetilde R)$, agent $4$ points to $h(4)$, agents $1$ and $2$ point to $h(3)$, and agent $3$ points to $h(1)$; hence, agent $4$ keeps his house, and agents $1$ and $3$ trade. At Step~2, agent $2$ is the only remaining agent and keeps his house. Thus, the trading cycles of $\varphi^\succ(\Nc,\widetilde R)$ are
\[
1\to h(3)\to 3\to h(1)\to 1,\quad 2\to h(2)\to 2,\quad\text{and}\quad 4\to h(4)\to 4,
\]
whereas the only trading cycle of $\widetilde\psi(\Nc,\widetilde R)=\widetilde\mu$ is
\[
1\to h(2)\to 2\to h(4)\to 4\to h(3)\to 3\to h(1)\to 1.
\]
The associated type allocations are
\[
\tau(\varphi^{\succ}(\Nc,\widetilde R))=(b,a,a,b)\quad\text{and}\quad\tau(\widetilde\mu)=(a,b,a,b);
\]
hence, agents $1$ and $2$ receive houses of different types at the two matchings, and the two matchings are not welfare equivalent.

\begin{lemma}[TTC outcomes at $\widetilde R^1$, $\widetilde R^4$, and $\widetilde R^{14}$]\label{lem:compA}\leavevmode
\begin{enumerate}[label=\textup{(\alph*)},nosep]
\item At $\varphi^\succ(\Nc,\widetilde R^1)$, agents $1$ and $4$ keep their houses, and agents $2$ and $3$ exchange their houses.
\item At $\varphi^\succ(\Nc,\widetilde R^4)$, agents $1$ and $3$ exchange their houses, and agents $2$ and $4$ exchange their houses.
\item $\varphi^\succ(\Nc,\widetilde R^{14})=\varphi^\succ(\Nc,\widetilde R^1)$.
\end{enumerate}
In particular, $\varphi^\succ(\Nc,\widetilde R^1)$ and $\varphi^\succ(\Nc,\widetilde R^{14})$ are welfare equivalent to $\widetilde\mu$, agent $1$ receives $h(1)$ at $\varphi^\succ(\Nc,\widetilde R^1)$, and agent $4$ receives a house of type $a$ at $\varphi^\succ(\Nc,\widetilde R^4)$.
\end{lemma}

\begin{proof}
(a) Given the strict preferences listed in Table~\ref{tab:Rtilde}, at Step~1 of the TTC algorithm for $\succ$ at $(\Nc,\widetilde R^1)$, agents $1$ and $4$ point to their own houses, agent $2$ points to $h(3)$, and agent $3$ points to $h(1)$; hence, agents $1$ and $4$ are removed with their own houses. At Step~2, agent $3$ points to $h(2)$ and agent $2$ still points to $h(3)$; hence, agents $2$ and $3$ trade. Thus, agents $1$ and $4$ keep their houses, and agents $2$ and $3$ exchange their houses.\smallskip

\noindent (b) Given the strict preferences listed in Table~\ref{tab:Rtilde}, at Step~1 of the TTC algorithm for $\succ$ at $(\Nc,\widetilde R^4)$, agents $3$ and $4$ point to $h(1)$ and agents $1$ and $2$ point to $h(3)$; hence, agents $1$ and $3$ trade. At Step~2, agent $4$ points to $h(2)$ and agent $2$ points to $h(4)$; hence, agents $2$ and $4$ trade. Thus, agents $1$ and $3$ exchange their houses, and agents $2$ and $4$ exchange their houses.\smallskip

\noindent (c) Given the strict preferences listed in Table~\ref{tab:Rtilde}, at Step~1 of the TTC algorithm for $\succ$ at $(\Nc,\widetilde R^{14})$, agent $1$ points to his own house, agents $3$ and $4$ point to $h(1)$, and agent $2$ points to $h(3)$; hence, the only cycle is $1\to h(1)\to 1$. At Step~2, agents $3$ and $4$ point to $h(2)$ and agent $2$ still points to $h(3)$; hence, agents $2$ and $3$ trade. At Step~3, agent $4$ is the only remaining agent and keeps his house. Thus, agents $1$ and $4$ keep their houses, and agents $2$ and $3$ exchange their houses, as at $\varphi^\succ(\Nc,\widetilde R^1)$.
\end{proof}

\begin{proposition}[A four-agent counterexample]\label{prop:four}
The rule $\widetilde\psi$ satisfies \textsl{individual rationality}, \textsl{Pareto efficiency}, \textsl{group strategy-proofness}, \textsl{non-bossiness}, and \textsl{consistency}. However, $\widetilde\psi$ is not welfare equivalent to any TTC rule with fixed tie-breaking.
\end{proposition}

\begin{proof}
By Lemma~\ref{lem:ttc}, $\varphi^\succ$ satisfies all five properties. Since $\widetilde\psi$ differs from $\varphi^\succ$ only at $(\Nc,\widetilde R)$, it suffices to check the requirements of the properties that involve the problem $(\Nc,\widetilde R)$.\smallskip

\noindent\emph{Individual rationality and Pareto efficiency.} Since, at $\widetilde\mu$, agent $1$ receives a house of his own type and agents $2$, $3$, and $4$ receive houses of their preferred types, $\widetilde\mu$ is \textsl{individually rational} for $\widetilde R$. Since, at $\widetilde\mu$, agents $2$, $3$, and $4$ receive houses of their preferred types, none of them can be made better off. Agent $1$ can only be made better off by receiving a house of type $b$. However, at $\widetilde\mu$, both houses of type $b$ are assigned to agents $2$ and $4$. Hence, if agent $1$ receives one of them, then agent $2$ or agent $4$ receives a house of type $a$ instead, and this agent is worse off. Thus, $\widetilde\mu$ is \textsl{Pareto efficient} for $\widetilde R$.\smallskip

\noindent\emph{Group strategy-proofness.} Suppose, by contradiction, that there exist a problem $(N,R)$, a group $S\subseteq N$, and $R'_S\in\Rd(H_N)^S$ such that $S$ manipulates $\widetilde\psi$ at $(N,R)$ via $R'_S$. Let $R'=(R'_S,R_{-S})$ and let $S'=\{i\in S:R'_i\neq R_i\}\subseteq S$ be the set of members of $S$ who change their preferences. Since some member of $S$ strictly gains, $R'\neq R$, i.e., $S'\neq\emptyset$. Since $\widetilde\psi$ coincides with $\varphi^\succ$ except at $(\Nc,\widetilde R)$ and $\varphi^\succ$ satisfies \textsl{group strategy-proofness}, $N=\Nc$ and $\widetilde R\in\{R,R'\}$, i.e., $\widetilde\mu=\widetilde\psi(\Nc,\widetilde R)$ is the outcome either at the true preferences $R$ or at the reported preferences $R'$. Since there are two types and preferences are strict across types, each agent $i\in S'$ ranks the two types at $R'_i$ in the opposite order from $R_i$.

First, suppose that $R=\widetilde R$, i.e., $\widetilde\psi(\Nc,R)=\widetilde\mu$. Since, at $\widetilde\mu$, agents $2$, $3$, and $4$ receive houses of their preferred types, only agent $1$ can strictly gain; hence, agent $1$ receives a house of type $b$ at $\widetilde\psi(\Nc,R')$. Since $R'\neq\widetilde R$, the definition of $\widetilde\psi$ implies that $\widetilde\psi(\Nc,R')=\varphi^\succ(\Nc,R')$. Since, by Lemma~\ref{lem:ttc}, $\varphi^\succ$ satisfies \textsl{individual rationality}, each agent receives at $\widetilde\psi(\Nc,R')$ a house that he weakly prefers to his own house according to his preferences at $R'$. If $1\in S'$, then agent $1$ prefers type $a$ at $R'$ and therefore receives a house of type $a$ at $\widetilde\psi(\Nc,R')$; hence, $1\notin S'$. If $i\in S'\cap\{2,3\}$, then agent $i$ prefers his own type at $R'$ and hence, at $\widetilde\psi(\Nc,R')$, receives a house of his own type, which he truly ranks below the house of the other type that he receives at $\widetilde\mu$. Since each member of $S$, and hence each member of $S'$, weakly gains, it follows that $S'\cap\{2,3\}=\emptyset$. Hence, $S'=\{4\}$, i.e., $R'=\widetilde R^4$. By Lemma~\ref{lem:compA}(b), at $\widetilde\psi(\Nc,\widetilde R^4)$, agent $4$ receives a house of type $a$, which he truly ranks below the house of type $b$ that he receives at $\widetilde\mu$, a contradiction.

Second, suppose that $R'=\widetilde R$, i.e., $\widetilde\psi(\Nc,R')=\widetilde\mu$. Then, each agent $i\in S'\cap\{2,3\}$ truly prefers his own type. Since $R\neq\widetilde R$, the definition of $\widetilde\psi$ implies that $\widetilde\psi(\Nc,R)=\varphi^\succ(\Nc,R)$. Since, by Lemma~\ref{lem:ttc}, $\varphi^\succ$ satisfies \textsl{individual rationality}, each such agent receives a house of his own type at $\widetilde\psi(\Nc,R)$ and a house of the other type at $\widetilde\mu$, and hence does not weakly gain. Since each member of $S$, and hence each member of $S'$, weakly gains, it follows that $S'\subseteq\{1,4\}$. If $S'=\{1\}$ or $S'=\{1,4\}$, then $R=\widetilde R^1$ or $R=\widetilde R^{14}$, and, by Lemma~\ref{lem:compA}(a) and (c), $\widetilde\psi(\Nc,R)$ is welfare equivalent to $\widetilde\mu$; hence, no agent strictly gains, a contradiction. If $S'=\{4\}$, then $R=\widetilde R^4$ and agent $4$ truly prefers type $a$; by Lemma~\ref{lem:compA}(b), agent $4$ receives a house of type $a$ at $\widetilde\psi(\Nc,\widetilde R^4)$ and a house of type $b$ at $\widetilde\mu$, and hence he does not weakly gain, a contradiction.\smallskip

\noindent\emph{Non-bossiness.} Since $\widetilde\psi$ coincides with $\varphi^{\succ}$ except at $(\Nc,\widetilde R)$ and $\varphi^{\succ}$ satisfies \textsl{non-bossiness}, it suffices to compare $(\Nc,\widetilde R)$ with the problems at which the preferences of a single agent differ from $\widetilde R$. Let $i\in\Nc$, let $R'_i\in\Rd(H_\Nc)$ with $R'_i\neq\widetilde R_i$, and let $R'=(R'_i,\widetilde R_{-i})$. We check \textsl{non-bossiness} for both changes, from $\widetilde R_i$ to $R'_i$ and from $R'_i$ to $\widetilde R_i$. Since $R'\neq\widetilde R$, the definition of $\widetilde\psi$ implies that $\widetilde\psi(\Nc,R')=\varphi^\succ(\Nc,R')$. If $i=1$, then, by Lemma~\ref{lem:compA}(a), agent $1$ receives $h(1)$ at $\widetilde\psi(\Nc,R')$ and $h(2)$ at $\widetilde\mu$. If $i\in\{2,3\}$, then agent $i$ prefers his own type at $R'$; since $\varphi^\succ$ satisfies \textsl{individual rationality}, agent $i$ receives a house of his own type at $\widetilde\psi(\Nc,R')$ and a house of the other type at $\widetilde\mu$. If $i=4$, then, by Lemma~\ref{lem:compA}(b), agent $4$ receives a house of type $a$ at $\widetilde\psi(\Nc,R')$ and $h(3)$ at $\widetilde\mu$. Hence, in each case agent $i$'s allotment changes, and \textsl{non-bossiness} imposes no requirement in either direction. Thus, $\widetilde\psi$ satisfies \textsl{non-bossiness}.\smallskip

\noindent\emph{Consistency.} Since a reduced problem has fewer agents than $\Nc$, the problem $(\Nc,\widetilde R)$ appears in the requirements of \textsl{consistency} only as the original problem, never as a reduced problem. Since $\widetilde\mu$ consists of the single trading cycle $1\to h(2)\to 2\to h(4)\to 4\to h(3)\to 3\to h(1)\to 1$, at $\widetilde\mu$, no non-empty group $S\subsetneq\Nc$ has allotments that are precisely its endowments, and \textsl{consistency} imposes no requirement at $(\Nc,\widetilde R)$. Hence, $\widetilde\psi$ satisfies \textsl{consistency}.\footnote{\citet[Section~5.4]{Ehlers2014} uses the same construction to show that \textsl{non-bossiness} cannot be dropped from his characterization: his rule differs from a TTC rule with fixed tie-breaking only at problems at which all agents trade in a single trading cycle, so that \textsl{consistency} imposes no requirement at these problems. At such a problem, an agent can report a strict preference that breaks a tie in his preferences; this report leaves his own allotment unchanged but changes the allotments of other agents. Hence, his rule violates \textsl{non-bossiness}. On the domain of objective indifferences, an agent cannot break a tie between identical houses, and $\widetilde\psi$ satisfies \textsl{non-bossiness}.}\smallskip

\noindent\emph{$\widetilde\psi$ is not welfare equivalent to a TTC rule with fixed tie-breaking.} Suppose, by contradiction, that for some tie-breaking structure $\succ'$, $\widetilde\psi$ is welfare equivalent to $\varphi^{\succ'}$. Consider the problem for coalition $\{1,2,3\}$ at which each agent prefers the other type to his own type. Since $h(3)$ is the only house of type $b$ in this coalition, agents $1$ and $2$ point to $h(3)$ at Step~1 of the TTC algorithm for $\succ'$; since agent $3$ points to whichever of $h(1)$ and $h(2)$ he ranks higher under $\succ'_3$, he trades with its owner. At this problem, $\widetilde\psi$ coincides with $\varphi^\succ$, under which agent $3$ trades with agent $1$; hence, $1\succ'_3 2$.

Now consider the TTC algorithm for $\succ'$ at $(\Nc,\widetilde R)$. Since agent $3$ prefers type $a$ and $1\succ'_3 2$, agent $3$ points to $h(1)$ as long as agent $1$ remains. Since $h(1)$ points to its owner, every cycle containing agent $3$ while agent $1$ remains must also contain agent $1$. Hence, agent $3$ cannot be removed before agent $1$, and $h(3)$ remains as long as agent $1$ remains. Since agent $1$ prefers type $b$ and $h(3)$ remains as long as agent $1$ remains, agent $1$ points to a house of type $b$ at every step at which he remains. Therefore, when agent $1$ is removed in a cycle, he receives a house of type $b$. However, agent $1$ receives a house of type $a$ at $\widetilde\psi(\Nc,\widetilde R)=\widetilde\mu$, a contradiction.
\end{proof}

The conclusion in Proposition~\ref{prop:four} that $\widetilde\psi$ is not welfare equivalent to any TTC rule with fixed tie-breaking uses the behavior of $\widetilde\psi$ on the proper subcoalition $\{1,2,3\}$. Indeed, when both rules are restricted to the problems with all four agents, $\widetilde\psi(\Nc,\cdot)$ is welfare equivalent to a TTC rule with fixed tie-breaking. To see this, let $\widehat\succ$ be obtained from $\succ$ only by reversing agent $3$'s ranking of the two houses of type $a$, so that $2\mathrel{\widehat\succ}_3 1$, while leaving all other same-type comparisons unchanged. Running the TTC algorithm at each of the $16$ preference profiles for $\Nc$ shows that this reversal leaves the type allocation unchanged at every preference profile for $\Nc$ other than $\widetilde R$; at $\widetilde R$, it switches the cross-type trade from agent $1$ to agent $2$ and yields the type allocation of $\widetilde\mu$. Hence, $\varphi^{\widehat\succ}(\Nc,\cdot)$ is welfare equivalent to $\widetilde\psi(\Nc,\cdot)$. Thus, Proposition~\ref{prop:four} does not contradict the fixed-population minimality statement in Remark~\ref{rem:min}.

\begin{remark}[Minimality on all coalitions]\label{rem:min4}
Four agents are the smallest number of agents for which a two-type rule as in Proposition~\ref{prop:four} exists, and, since with one type all matchings are welfare equivalent, two types are minimal. It remains to rule out a two-type counterexample with at most three agents. Up to relabeling of the types, the only non-trivial type composition is $aab$. Let agents $1$ and $2$ own houses of type $a$, let agent $3$ own a house of type $b$, and let $f$ satisfy \textsl{individual rationality} and \textsl{Pareto efficiency}. These two properties determine every agent's welfare at every problem except the problem with all three agents at which each of them prefers the other type. At that problem, agent $3$ must trade with exactly one of agents $1$ and $2$, while the other owner of a house of type $a$ receives a house of type $a$. Choose a tie-breaking structure $\succ'$ in which agent $3$ ranks the house of that trading partner above the other house of type $a$. Then $f$ and $\varphi^{\succ'}$ are welfare equivalent. Hence, no two-type counterexample exists with fewer than four agents.
\end{remark}

\section{Minimality for a fixed population}\label{app:search}

In this appendix, we establish the fixed-population minimality statement in Remark~\ref{rem:min} by a computer search. The two Python scripts used, \texttt{Claude\_minimality\_check.py} and \texttt{ChatGPT\_minimality\_check.py}, are available, each with a README file, as ancillary files of the arXiv version of this paper. Recall that Theorem~\ref{thm:main} provides, for a population with type composition $aaabb$, a rule $f$ (namely, $\psi$) that satisfies \textsl{individual rationality}, \textsl{Pareto efficiency}, and \textsl{group strategy-proofness} (as well as \textsl{non-bossiness} and \textsl{consistency}), but whose restriction $f(\Nc,\cdot)$ to the problems with all agents is not welfare equivalent to $\varphi^\succ(\Nc,\cdot)$ for any tie-breaking structure $\succ$. Remark~\ref{rem:min} states that, with two types of houses, five agents are the smallest number of agents for which such a fixed-population counterexample exists. To establish this, we show the following, stronger statement: for each two-type population $\Nc$ with at most five agents and another type composition, i.e., with type composition $aab$, $aaab$, $aabb$, or $aaaab$ (up to relabeling of the types), the restriction $f(\Nc,\cdot)$ of each rule $f$ that satisfies \textsl{individual rationality}, \textsl{Pareto efficiency}, and \textsl{group strategy-proofness} is welfare equivalent to $\varphi^\succ(\Nc,\cdot)$ for some tie-breaking structure $\succ$. Since each rule that satisfies the five properties of Theorem~\ref{thm:main} in particular satisfies \textsl{individual rationality}, \textsl{Pareto efficiency}, and \textsl{group strategy-proofness}, this excludes a fixed-population counterexample as in Theorem~\ref{thm:main} for each of these populations.\medskip

Throughout this appendix, let $\Nc$ be a population whose houses have two types, and consider only problems $(\Nc,R)$ with all agents in $\Nc$. Since there are two types and preferences are strict across types, each agent has two possible preferences, and there are $2^{|\Nc|}$ preference profiles.\medskip

A type allocation $t=(t_i)_{i\in\Nc}$ specifies for each agent $i\in\Nc$ the type $t_i\in\T$ of the house he receives. It is \emph{feasible} if it is induced by some matching, i.e., if $t=\tau(\mu)$ for some matching $\mu$ for $\Nc$; equivalently, for each type, the number of agents who receive a house of that type equals the number of houses of that type. Since welfare depends only on types, the restriction $f(\Nc,\cdot)$ of a rule $f$ is, up to welfare equivalence, described by the function $W_f$ that assigns to each preference profile $R$ the feasible type allocation $W_f(R)=\tau(f(\Nc,R))$. More generally, we call a function $W$ that assigns to each preference profile a feasible type allocation a \emph{welfare rule}. \textsl{Individual rationality}, \textsl{Pareto efficiency}, and \textsl{group strategy-proofness} are defined for welfare rules as for rules, with each agent's allotment replaced by its type. Let $\mathcal{W}$ denote the set of welfare rules that satisfy \textsl{individual rationality}, \textsl{Pareto efficiency}, and \textsl{group strategy-proofness}, and let $\mathcal{W}^{T\!T\!C}$ denote the set of welfare rules $W_{\varphi^\succ}$ induced by the TTC rules $\varphi^\succ$ based on tie-breaking structures $\succ$ in which each agent ranks himself first. By Lemma~\ref{lem:ttc}, $\mathcal{W}^{T\!T\!C}\subseteq\mathcal{W}$. If $\mathcal{W}^{T\!T\!C}=\mathcal{W}$, then for each rule $f$ that satisfies \textsl{individual rationality}, \textsl{Pareto efficiency}, and \textsl{group strategy-proofness}, $f(\Nc,\cdot)$ is welfare equivalent to $\varphi^\succ(\Nc,\cdot)$ for some tie-breaking structure $\succ$, and hence no fixed-population counterexample as in Theorem~\ref{thm:main} exists for $\Nc$. We determine $\mathcal{W}$ and $\mathcal{W}^{T\!T\!C}$ with a Python script written with the assistance of Claude (Anthropic), \texttt{Claude\_minimality\_check.py}, which performs the three computations (i)--(iii) described next.\medskip

\noindent\textup{(i)} \emph{Individual rationality and Pareto efficiency.} For each preference profile $R$, the Python script determines, by checking all feasible type allocations, those feasible type allocations that are \textsl{individually rational} and \textsl{Pareto efficient} for $R$; we call them the \emph{candidates} at $R$. Each welfare rule in $\mathcal{W}$, and hence each welfare rule in $\mathcal{W}^{T\!T\!C}$, selects a candidate at each preference profile.\medskip

\noindent\textup{(ii)} \emph{Group strategy-proofness.} The Python script determines $\mathcal{W}$ by solving a system of logical constraints. For each preference profile $R$ and each candidate $t$ at $R$, it introduces a variable $x_{R,t}\in\{0,1\}$, with the interpretation that $x_{R,t}=1$ if and only if the welfare rule selects $t$ at $R$. First, for each preference profile $R$, exactly one of the variables $x_{R,t}$ equals $1$; hence, each assignment of values to the variables that satisfies these constraints describes a welfare rule that selects a candidate at each preference profile. Second, recall that a group $S\subseteq\Nc$ manipulates a welfare rule at $R$ via $R'_S$ if, for $R'=(R'_S,R_{-S})$, each member of $S$ weakly prefers, according to $R$, the type he receives at $R'$ to the type he receives at $R$, and some member of $S$ strictly prefers it. Let $S'=\{i\in S:R'_i\neq R_i\}\subseteq S$ be the set of members of $S$ who change their preferences; equivalently, $S'$ is the set of agents whose preferences differ at $R$ and $R'$. Hence, a welfare rule violates \textsl{group strategy-proofness} at $R$ if and only if there is a preference profile $R'\neq R$ such that, according to $R$, each agent in $S'$ weakly prefers the type he receives at $R'$ to the type he receives at $R$ and some agent $j\in\Nc$ strictly prefers it: if such $R'$ exists, the group $S=S'\cup\{j\}$ manipulates the welfare rule at $R$ via $R'_S$, where, if $j\notin S'$, agent $j$ reports his true preferences; conversely, if some group $S$ manipulates the welfare rule at $R$ via $R'_S$, then $S'\subseteq S$, and the condition holds for $R'=(R'_S,R_{-S})$. Accordingly, for each pair of distinct preference profiles $R$ and $R'$, with $S'=\{i\in\Nc:R'_i\neq R_i\}$, each candidate $t$ at $R$, and each candidate $t'$ at $R'$ such that, according to $R$, each agent $i\in S'$ weakly prefers type $t'_i$ to type $t_i$ and some agent $j\in\Nc$ strictly prefers type $t'_j$ to type $t_j$, we require that $x_{R,t}+x_{R',t'}\leq 1$, i.e., that the welfare rule does not select both $t$ at $R$ and $t'$ at $R'$. Since the first constraints ensure that an assignment describes a welfare rule that selects a candidate, i.e., an \textsl{individually rational} and \textsl{Pareto efficient} type allocation, at each preference profile, and the second constraints exclude exactly the violations of \textsl{group strategy-proofness}, the assignments that satisfy all constraints correspond exactly to the welfare rules in $\mathcal{W}$. The problem of finding an assignment of values $0$ or $1$ to variables that satisfies such logical constraints is known as a satisfiability (SAT) problem, and standard computer programs, called SAT solvers, solve it. The Python script uses the SAT solver CaDiCaL \citep{BiereEtAl2024}, version 1.5.3, which it calls through the Python package PySAT \citep{IgnatievMorgadoMarquesSilva2018}. To obtain all welfare rules in $\mathcal{W}$, the Python script solves the system repeatedly: after each solution, it adds a constraint that excludes exactly this solution and solves again, until no further solution exists.\medskip

\noindent\textup{(iii)} \emph{TTC rules with fixed tie-breaking.} The Python script determines $\mathcal{W}^{T\!T\!C}$ as follows. For each tie-breaking structure $\succ$ in which each agent ranks himself first, it runs the TTC algorithm at each preference profile $R$ and records the welfare rule $W_{\varphi^\succ}$. Since $\varphi^\succ$ depends on $\succ$ only through each agent's rankings of houses of the same type, and since an agent who prefers his own type and ranks himself first points to his own house, only each agent's ranking of the owners of houses of the other type matters. Table~\ref{tab:search} reports the number of such tie-breaking structures.

\begin{table}[ht]
\centering
\begin{tabular}{lcccc}
\hline
Type & Preference & Tie-breaking & & \\
composition & profiles & structures & $|\mathcal{W}^{T\!T\!C}|$ & $|\mathcal{W}|$ \\
\hline
$aab$ & $8$ & $2$ & $2$ & $2$ \\
$aaab$ & $16$ & $6$ & $6$ & $6$ \\
$aabb$ & $16$ & $16$ & $16$ & $16$ \\
$aaaab$ & $32$ & $24$ & $24$ & $24$ \\
$aaabb$ & $32$ & $288$ & $288$ & $864$ \\
\hline
\end{tabular}
\caption{Numbers of preference profiles, tie-breaking structures in which each agent ranks himself first, welfare rules in $\mathcal{W}^{T\!T\!C}$ (induced by the corresponding TTC rules), and welfare rules in $\mathcal{W}$ (satisfying \textsl{individual rationality}, \textsl{Pareto efficiency}, and \textsl{group strategy-proofness}).}\label{tab:search}
\end{table}
\FloatBarrier

One-type populations are welfare-trivial, and a population with type composition $ab$ is a strict-preference problem. Up to relabeling of the types, the type compositions of the remaining two-type populations with at most five agents are those in Table~\ref{tab:search}. For populations with type composition $aab$, $aaab$, $aabb$, or $aaaab$, $\mathcal{W}^{T\!T\!C}=\mathcal{W}$. Hence, for these populations, the welfare rules satisfying \textsl{individual rationality}, \textsl{Pareto efficiency}, and \textsl{group strategy-proofness} are exactly the welfare rules induced by TTC rules with fixed tie-breaking, and no fixed-population counterexample exists. Together with Theorem~\ref{thm:main}, this establishes the fixed-population minimality statement in Remark~\ref{rem:min}.

For the population with type composition $aaabb$, the Python script also runs the TTC algorithm for all $12^5=248{,}832$ tie-breaking structures, i.e., for each agent, all rankings of the owners of houses of type $a$ and of the owners of houses of type $b$, including those in which some agent does not rank himself first. These tie-breaking structures induce $336$ distinct welfare rules, which, by Lemma~\ref{lem:ttc}, all belong to $\mathcal{W}$. Hence, $528$ of the $864$ welfare rules in $\mathcal{W}$ are not induced by any TTC rule with fixed tie-breaking. As a cross-check of Theorem~\ref{thm:main}, the welfare rule $W_\psi$ induced by the rule $\psi$ of Theorem~\ref{thm:main} is one of them. The minimality computations were independently checked with a separate Python script written with the assistance of ChatGPT (OpenAI), \texttt{ChatGPT\_minimality\_check.py}, which is based on exhaustive enumeration and backtracking rather than a SAT solver.

\providecommand{\enquote}[1]{``#1''}
\begin{spacing}{1.20}

\end{spacing}

\end{document}